\documentclass[fleqn,usenatbib]{mnras}

\usepackage{newtxtext,newtxmath}
\usepackage{comment}
 
\usepackage[T1]{fontenc}

\DeclareRobustCommand{\VAN}[3]{#2}
\let\VANthebibliography\thebibliography
\def\thebibliography{\DeclareRobustCommand{\VAN}[3]{##3}\VANthebibliography}
\def\v{\upsilon}

\usepackage{graphicx}	
\usepackage{amsmath}	
\usepackage{amssymb}	

\title[Silicate/volatile disc model around white dwarfs]{Modeling the Evolution of Silicate/Volatile Accretion Discs around White Dwarfs}

\author[A. Okuya et al.]{
Ayaka Okuya,$^{1}$\thanks{E-mail: ayaka.okuya@nao.ac.jp}
Shigeru Ida,$^{2}$
Ryuki Hyodo,$^{3}$
and Satoshi Okuzumi$^{4}$
\\
$^{1}$National Astronomical Observatory of Japan, 2-21-1 Osawa, Mitaka, Tokyo 181-8588, Japan\\
$^{2}$Earth-Life Science Institute, Tokyo Institute of Technology, Ookayama, Meguro-ku, Tokyo 152-8550, Japan\\
$^{3}$ISAS/JAXA, Sagamihara, Kanagawa, Japan \\
$^{4}$Department of Earth and Planetary Sciences, Tokyo Institute of Technology, Ookayama, Meguro-ku, Tokyo 152-8551, Japan
}

\date{Accepted 2022 November 27. Received 2022 November 7; in original form 2021 August 31}

\pubyear{2022}

\begin{document}
\label{firstpage}
\pagerange{\pageref{firstpage}--\pageref{lastpage}}
\maketitle

\begin{abstract}
A growing number of debris discs have been detected around metal-polluted white dwarfs. They are thought to be originated from tidally disrupted exoplanetary bodies and responsible for metal accretion onto host WDs. 
To explain (1) the observationally inferred accretion rate higher than that induced by Poynting-Robertson drag, $\dot{M}_{\rm PR}$, and (2) refractory-rich photosphere composition indicating the accretion of terrestrial rocky materials, previous studies proposed runaway accretion of silicate particles due to gas drag by the increasing silicate vapor produced by the sublimation of the particles. 
Because re-condensation of the vapor diffused beyond the sublimation line was neglected, we revisit this problem by one-dimensional advection/diffusion simulation that consistently incorporates silicate sublimation/condensation and back-reaction to particle drift due to gas drag in the solid-rich disc. 
We find that the silicate vapor density in the region overlapping the solid particles follows the saturating vapor pressure and that no runaway accretion occurs if the re-condensation is included.
This always limits the accretion rate from mono-compositional silicate discs to $\dot{M}_{\rm PR}$ in the equilibrium state.
Alternatively, by performing additional simulations that couple the volatile gas (e.g., water vapor),
we demonstrate that the volatile gas enhances the silicate
accretion to $>\dot{M}_{\rm PR}$ through gas drag.
The refractory-rich accretion is simultaneously reproduced when the initial volatile fraction of disc is $\la 10$ wt\% because of the suppression of volatile accretion due to the efficient back-reaction of solid to gas.
The discs originating from C-type asteroid analogs might be a possible clue to the high-$\dot{M}$ puzzle.


\end{abstract}




\section{Introduction}

A growing number of observations show the universal presence of planetary materials in and around white dwarfs (WDs).
From a quarter to a half of WDs have metals such as Fe, Si, O, and C in their atmospheres \citep[e.g.,][]{Zuckerman+2003,Zuckerman+2010,Koester+2014,Hollands+2017}.
Some of them show infrared excess from their circumstellar dusty discs  \citep{Zuckerman+1987,Jura+2007, Rocchetto+2015}, and a few of them exhibit double-peaked emission lines \citep{Gansicke+2006, Gansicke+2007, Melis+2010, Manser+2016} or Doppler-broadened absorption lines \citep{Xu+2016, Steele+2021} from the circumstellar gas. These are thought to be originated from the remnant planetary systems around WDs which evolve with their host stars' post-main sequence evolution. 
The surviving asteroids and/or minor planets could be gravitationally scattered into star-grazing orbits \citep{Debes+2002,Bonsor+2011,Mustill+2018}, and the tidal disruption and subsequent collisions of the fragments generated by the disruption is likely to eventually produce compact dusty accretion discs \citep[e.g.,][]{Jura2003}.

Based on this scenario, an accretion disc can play a crucial role to link such observations to accreted bodies as well as their origin.
Moreover, to reconcile the short sinking timescale of heavier elements in the WD atmosphere due to the strong gravitationally field \citep{Paquette1986, Koester2010} with the universal metal detection, the continuous accretion of heavy elements from the circumstellar disc would be more likely rather than intermittent accretion such as the direct hitting of planetary bodies onto WDs.

The observed spectral energy distributions \citep{Jura+2007, Jura+2009} show that
the dust emission from the disc extends from 10-40 stellar radius  
out to the Roche limit radius ($r_{\rm R}$).
The outer edge at $\sim r_{\rm R}$ is consistent with a tidally disrupted disc.
The inner disc radius may correspond to the silicate sublimation line, suggesting that 
the particles\footnote{In this paper, we refer to silicate dust/solid particles
as {\it dust, particles, solid}, or their combinations, 
and to gaseous components produced by sublimation of solid particles
as {\it gas} or {\it vapor}.
The different words do not represent different meanings, but reflect the context.
For example, {\it solid} and {\it vapor} when considering phase change through sublimation/condensation, {\it gas} and {\it dust/particles} 
when describing their physical dynamics,  
and {\it dust} is also used when concerned with observations.
Following the conventional notation, we also adopt the abbreviations ``d'' (dust) and ``g'' (gas) for the subscripts for physical quantities.} in the disc are silicate debris particles of $\mu$m--cm sizes \citep{Graham1990,Jura+2009} and 
a silicate vapor disc would be connected with the particle disc at its inner edge 
and be distributed toward the stellar surface \citep{Rafikov2011b, Metzger+2012}.
The connected silicate vapor disc is observationally supported
by the detection of metal emission lines \citep[e.g.,][]{Gansicke+2006,Manser+2016}.
The observations also show that the vapor and the particles spatially overlap \citep{Brinkworth+2009, Melis+2010}.

The observationally inferred disc accretion rate onto the host star is $10^6 - 10^{11}$ g/s \citep{Farihi+2009,Farihi+2010, Chen+2019}.
Because the outer disc is likely to consist of solid debris particle
and it provides accreting vapor to the inner disc, 
it is important to clarify the inward transport mechanism of the debris.
The Poynting Robertson (PR) drag is a promising mechanism in the strong WD radiation field
\citep{Rafikov2011,Bochkarev+2011}.
However, as the accretion rate driven by PR-drag increases, the disc becomes optically thick
and the WD radiation received by individual solid particles deceases.
As a result, the accretion rate does not increase over $\dot{M}_{\rm PR,thick}\sim 10^8$ g/s,
once the disc becomes optically thick 
\citep[][see also section \ref{sec:model}]{Rafikov2011}, and therefore the higher observed accretion rate of $10^8 - 10^{11}$ g/s, which WDs with discs typically have \citep{Jura+2007,Farihi+2009}, cannot be explained.

The observed metal abundance in the WD atmospheres so far suggests that the vast majority of WDs accrete materials with terrestrial rocky composition:
Rock-forming refractories such as Fe, Mg, Si, and O
are commonly detected, and they comprise at least 85\% by mass of polluting materials \citep{Jura&Young2014}.
On the other hand, the detection of volatile elements such as C, N, and S are currently very rare \citep{Xu+2017}.
The deficient pattern of these elements resembles dry rocky bodies in the solar systems \citep{Gansicke+2012,Jura&Young2014}.
In addition, Ca and other refractories are detected in the gas disc emission line in several systems whereas volatiles such as H and O has been observed in only one system \citep{Gentile-Fusillo+2021}.

While a few ideas are suggested to account for the higher observed accretion rates than $\dot{M}_{\rm PR,thick}$
\citep{Jura2008,Farihi+2012}, 
\citet{Rafikov2011b} and \citet{Metzger+2012} proposed a detailed idea in the framework of debris particle drift by PR-drag that 
the silicate particle drift is accelerated by
gas drag from the silicate vapor produced by sublimation of the particle at the location with the silicate sublimation temperature (the silicate line). 
The produced vapor viscously spreads outward to interact with the particles beyond the silicate line. 
Because of the differential rotation between vapor and particles due to radial pressure gradient in the vapor disc,
the particles lose angular momentum by gas drag \citep{Adachi+1976}.
The acceleration of the particle drift enhances the sublimation and therefore the gas drag.
By this positive feedback, the drift by gas drag becomes faster than that by PR-drag, and the accretion rate increases in a runaway fashion until the particles are completely depleted
\citep[][see Appendix \ref{app:M12} for the detailed model description]{Metzger+2012}.
\citet{Metzger+2012} found that their simulations with the runaway accretion successfully account for (1) the higher accretion rate that cannot be obtained by PR drag alone and (2) the photospheric composition polluted by rocky materials.

The essential point of the runaway accretion model by \citet{Metzger+2012} is 
the co-existence of the silicate vapor and particles.
They discussed the potential importance of re-condensation of the vapor
diffused outward because the characteristic condensation timescale may be short enough
to suppress the runaway accretion.
However, because the overlapping of the vapor and the particle distributions
is suggested by the observations,
\citet{Metzger+2012} neglected the vapor re-condensation.

In this paper, we revisit the high disc accretion rate onto WDs
by 1D advection/diffusion simulation
with more sophisticate sublimation/condensation calculation using the phase change rate as a function of vapor and the saturation pressures, following \citet{Hyodo+2019, Hyodo+2021}.
Because the vapor can exist only up to the saturation pressure beyond the silicate line, we show in this paper that
the silicate vapor and particle distributions hardly overlap at the disc mid-plane and that the runaway accretion is significantly suppressed in the case of a single component (silicate), as \citet{Metzger+2012} pointed out its possibility.
As the condensation effect is the most significant for the accretion rate, the runaway accretion would be inhibited independently of the detailed treatments of solid-gas interaction (see Sec. \ref{subsec:with-con}).

However, the possibility remains that the acceleration of 
the silicate particle drift is induced by the vapor of more volatile materials.
Dynamical arguments allow the supply of solid bodies from the wide range of orbital radii, such as Main Asteroid Belt analogs to Kuiper-Belt analogs \citep{Mustill+2018, Li+2022}.
This means that, although volatile detection has been currently rare in the WD atmospheres and/or disc emission lines, the delivery of ice-bearing bodies could be quite possible and they would form volatile-bearing accretion discs.
The Main Asteroid Belt analog may supply not only the dry S-type asteroids but also C-type asteroids which marginally include water of $\sim 10$ wt\%, into the Roche limit radius.
Because of a longer Keplerian timescale for a larger orbital radius, Kuiper-Belt analogs in outer regions can be massive reservoirs of the source of the pollution \citep{Bonsor+2010}.
Although the probability that the planetary bodies of an outer system can be gravitationally scattered onto star-grazing orbits highly depends on the distribution of an inner planetary system \citep{Bonsor+2011}, the infall of Kuiper-Belt analogs would deliver the approximately same amount of water as the rock components.

Accordingly, we also perform additional simulations that include volatile vapor (e.g., water vapor) produced by icy components of the debris. 
For the circumstellar discs around WDs, the snow line would be much farther than the Roche limit radius (see Figure \ref{fig:WD-disc-image}).
The icy components would be sublimated immediately after tidal capture/disruption. 
Because the water vapor does not condense in the region overlapping the silicate particles, it enhances the silicate accretion through gas drag, naturally covering the observational range of the accretion rates, for example, by changing the water vapor density.

The accretion from volatile-including discs seems to contradict the observed photospheric composition. 
However, the back-reaction forces of dust to gas can potentially mitigate the contradiction.
While silicate particles lose their angular momentum due to gas drag to drift to the central star, volatile vapor receives angular momentum from the solid particles as a back reaction and moves outward \citep[e.g.,][]{Nakagawa+1986, Kanagawa+2017}.
In addition, the volatile vapor beyond the snow line would condense into icy particles. If the drift of icy particles is slow, they may remain as a solid icy ring and become a reservoir of infalling volatile materials.
We investigate the possibility that volatile-bearing discs can produce not only a high accretion rate but also accretion mass with silicate-rich compositions through these processes of disc evolution.

The organization of this paper is as follows. 
In Section \ref{sec:model}, we describe the silicate disc model that solves the coupled evolution of silicate dust and vapor discs and our numerical settings. 
In Section \ref{sec:res-1comp}, we present the evolution and accretion rates of one-component silicate discs to identify the effects of condensation.
In Section \ref{sec:evo-w-vol}, we clarify the impacts of volatile vapor on the silicate disc evolution and demonstrate the possible condition for reproducing a higher range of observational rates and silicate-rich photospheric composition.
In Section \ref{sec:discussion}, we compare our simulation results with gas disc observations and elemental fraction observed in the photospheres to discuss their consistency.
We summarize our findings in Section \ref{sec:conclusion}.

\section{One-component disc Model} \label{sec:model}
\subsection{Key Features of Accretion Discs around White Dwarfs \label{subsec:major-feature}}

\begin{figure*}
\begin{center}
\includegraphics[bb= 0 0 482 223, width=17cm]{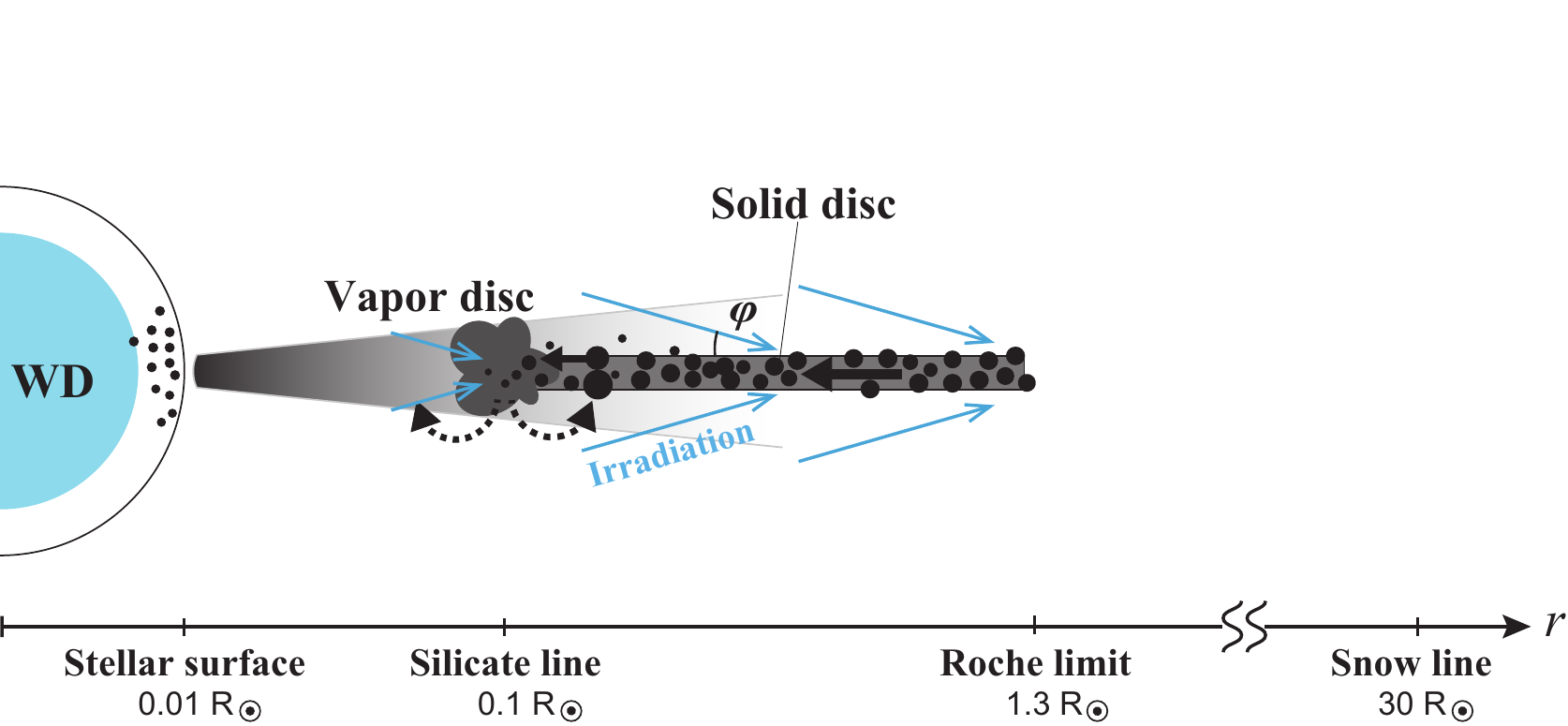}
\caption{A schematic showing the geometry of the modeled silicate particulate disc and the vapor discs around a WD. 
The incident angle of the radiation on the disc surface at distance $r$ from the WD is represented by $\phi$.}
\label{fig:WD-disc-image}
\end{center}
\end{figure*}

The silicate particle disc around a WD would be a compact disc extending from the silicate sublimation line to the Roche limit. 
Figure~\ref{fig:WD-disc-image} schematically shows the geometry of the modeled disc.
For materials with an internal density of $\rho_{\rm int}$, orbiting a
WD of physical radius $R_{\rm WD}$, mass $M_{\rm WD}$, and mean density $\rho_{\rm WD}$, 
the Roche limit radius is given by
\begin{align}
r_{\rm R} &= C_{\rm R} \left( \frac{\rho_{\rm WD}}{\rho_{\rm int}} \right)^{1/3} R_{\rm WD}\\ 
               & \sim 1.3 \ R_{\odot} \, \biggl(\frac{M_{\rm WD}}{0.6 M_{\odot}} \biggr)^{1/3} \biggl(\frac{\rho_{\rm int}}{3\ {\rm g\ cm^{-3}}}\biggr)^{-1/3}, \label{eq:roche}
\end{align}
where we assume $C_{\rm R}=2.0$ \citep{Lieshout+2018} and 
adopt $\rho_{\rm int}=$ 3 g\ cm$^{-3}$ for rocky materials.
We adopt $M_{\rm WD}=0.6M_{\odot}$ and $R_{\rm WD}=0.01\ R_{\odot}$, which are the typical values for the metal-polluted WDs.

As the silicate particles approach the sublimation line, silicate vapor is produced by the particle sublimation 
without background gas such as H and He.
We solve the sublimation and condensation rates of silicates based on 
the saturation pressure ($P_{\rm eq}$) and the local vapor pressure ($P_{\rm g}$) in the disc (see sec. \ref{subsec:g-d-change}).
As sublimation proceeds, the sublimation line radius ($r_{\rm sil}$), at which $P_{\rm eq} = P_{\rm g}$, evolves. 
Typically, $r_{\rm sil} \sim 0.1 \, R_{\odot}$ (see sec \ref{subsec:with-con}), where the disc temperature is 
$\sim 2000\, \rm K$ (Eq.~\ref{eq:T-prof}). 
This is considerably higher than the typical sublimation temperature used for silicates in protoplanetary discs, because in the disc around a WD that consists only of silicate vapor,
the vapor pressure is much higher than the partial pressure of silicates in the protoplanetary discs.
Such high sublimation temperature is consistent with the estimate by \citet{Rafikov2012} based on vapor pressure.

In this paper, we adopt the temperature determined by direct irradiation from the WD
with the photosphere temperature of 10000 K as \citep{Rafikov2012}
\begin{align}
T = 620\ 
 \biggl( \frac{R_{\rm WD}}{0.01 R_{\rm \odot}} \biggr)^{1/2}
 \biggl( \frac{T_{\rm WD}}{10000~{\rm K}} \biggr)\ 
\biggl( \frac{r}{r_{\rm R}} \biggr)^{-1/2} {\rm K}. \label{eq:T-prof}
\end{align}
Because the luminosity of WDs is high, the contribution of the viscous heating is negligible
compared to the stellar irradiation.
Adopting the isothermal sound speed,
$c_{\rm s} = \sqrt{k_{\rm B}T/\mu_{\rm g}}$ 
with the mean molecular weight $\mu_{\rm g} = 30\ m_{\rm p}$ corresponding to
Mg-SiO gas, where $k_{\rm B}$ is the Boltzmann constant and $m_{\rm p}$ is the proton mass,
the gas scale height for Eq.~(\ref{eq:T-prof}) is given by 
\begin{equation}
H_{\rm g} = c_{\rm s} \Omega^{-1} \sim  1.3 \times 10^{-3} \biggl(\frac{r}{1 R_{\odot}}\biggr)^{1/4} r, \label{eq:Hgas}
\end{equation}
where $\Omega$ is the Keplerian orbital frequency.
The disc is geometrically very thin;  
$H_{\rm g}$ at the Roche limit of $\sim 1 R_{\odot}$ is one tenth of the WD stellar radius, 
and the dust scale height is smaller than $H_{\rm g}$ (see Eq.~(\ref{eq:Hd})). 
For this disc geometry, the surface of an optically thick region is illuminated by the host star 
at an incident angle of \citep{Friedjung1985},
\begin{align}
\phi(r)=\frac{4}{3\pi}\frac{R_{\rm WD}}{r},
\label{eq:phi}
\end{align}
which is used for the calculation of Poynting-Roberton drag \citep[e.g.,][see also sec \ref{subsec:d-drift}]{Rafikov2011}.

The disc which formed from solid minor planets would initially have the gas surface density significantly lower than the dust particle surface density.
In such particle-dominated environments, particle-gas interaction such as gas drag is affected by surrounding other particles.
\citet{Metzger+2012} approximated the midplane particle layer as a solid plate and modeled the gas-drag force as a force exerted between the solid plate surface and the layer of gas above and below it.
On the other hand, our model assumes that each particle experience the gas drag from the gas phase individually but takes into account the particle collective effects by including the back-reaction of dust.
They are based on different physics approaches, but we show that mass accretion rate predicted by each gas-drag formulation possibly shares a similar dependence (Sec.~\ref{subsec:d-drift}), resulting in the approximately same disc evolution (Sec.~\ref{sec:res-1comp}).
Furthermore, the particle-based approach enables us to self-consistently incorporate the sublimation/condensation calculation into advection/diffusion model  (Sec.~\ref{subsec:g-d-change}.)

\subsection{Sublimation and Condensation of Silicates \label{subsec:g-d-change}}

\citet{Hyodo+2019, Hyodo+2021} simulated radially one-dimensional advection and diffusion of pebbles of ice-rock mixture
with sublimation/re-condensation of water ices near the snow line 
in a turbulent H-He gas protoplanetary disc
to discuss pile-ups of the silicate dust particles released from the pebbles
inside the snow line and the pebbles outside the snow line 
for the formation of rocky and icy planetesimals.
We follow their simulation methods, but with no background H-He gas, replacing the icy pebbles near the snow line by the silicate particles near the silicate line,
and adding the drift of the silicate particles due to PR-drag.
As we already mentioned, including re-condensation of silicate vapor
in a self-consistent way is the most important difference from previous 
studies on the WD disc evolution \citep{Bochkarev+2011, Metzger+2012}.

We follow \citet{S&O2017} and \citet{Hyodo+2019} for prescriptions of 
sublimation/re-condensation of silicates.
Macroscopic sublimation/re-condensation is regulated by adsorption of molecules in the gas phase to solid silicate particles 
and desorption of molecules from the particles to gas.
The former and latter rates are respectively represented by the partial vapor pressure $P_{\rm g}^{\rm sil}$
and saturation pressure $P_{\rm eq}$.
The local (partial) pressure of silicate vapor with surface density $\Sigma_{\rm g}^{\rm sil}$ and temperature $T$ is given by
\begin{align}
P_{\rm g}^{\rm sil} = \frac{\Sigma_{\rm g}^{\rm sil}}{\sqrt{2\pi} H_{\rm g}} \frac{k_{\rm B}T}{\mu_{\rm sil}},
\end{align}
where $\mu_{\rm sil} = 30\ m_{\rm p}$ for Mg-SiO gas.
The saturation pressure is given by the Clausius-Clapeyron equation as
\begin{align}
P_{\rm eq} = \exp(-\mathcal{A}/T + \mathcal{B})\ {\rm g\ s^{-1}\ cm^{-2}}. \label{eq:P-sat}
\end{align}
Assuming crystalline foresterite (Mg$_2$SiO$_4$) for a representative of silicate, we take values of $\mathcal{A}= 65309\ [{\rm K}]$ and $\mathcal{B}= 34.1$ \citep{Nagahara+1994}.

The mass change rate of a spherical solid particle with mass $m_{\rm d}$ 
is given by \citep{Lichtenegger+1991, Ciesla+2006, Rafikov2012}
\begin{align}
    \frac{dm_{\rm d}}{dt} =4 \pi r_{\rm d}^2 \v_{\rm th} \rho_{\rm d} \left(1- \frac{P_{\rm eq}}{P_{\rm g}^{\rm sil}}\right), \label{eq:m-change-rate}
\end{align}
where $r_{\rm d}$ is the physical radius of the particle, 
$\v_{\rm th}=\sqrt{8/\pi} \ c_{s}$ is the mean thermal velocity, and $\rho_{\rm d}$ is the spatial density
of silicate dust particles.

The location at $P_{\rm eq} = P_{\rm g}^{\rm sil}$ is called the sublimation line (the silicate line in our case).
Because $T$ decreases with increasing $r$ and $P_{\rm eq}$ is sensitive to $T$,
the dependence of the silicate line radius on $P_{\rm g}^{\rm sil}$ is weak. 
The vapor, diffused out to the region beyond the silicate line where $P_{\rm eq} < P_{\rm g}^{\rm sil}$,
re-condenses with the rate corresponding to Eq.~(\ref{eq:m-change-rate}) as follows.
The surface density change rate of solid particles due to sublimation/re-condensation,
which will be used as the source/sink term in Eq.~(\ref{eq:sigma-d}),
is given by
\begin{align}
\dot{\Sigma}_{\rm d} = (R_{\rm con}\Sigma_{\rm g}^{\rm sil} - R_{\rm eva}) {\Sigma}_{\rm d} \label{eq:dust-ss-term},
\end{align}
where $R_{\rm con}$ and  $R_{\rm eva}$ are
\begin{align}
R_{\rm con} & = 2\sqrt{\frac{k_{\rm B}T}{\mu_{\rm sil}}} \frac{r^{2}_{\rm d}}{m_{\rm d} H_{\rm g}}, \\
R_{\rm eva} & = 2\sqrt{2\pi} \frac{r^{2}_{\rm d}}{m_{\rm d}} \sqrt{\frac{\mu_{\rm sil}}{k_{\rm B}T}} P_{\rm eq}.
\end{align}
The vapor surface density change rate is $\dot{\Sigma}_{\rm g}^{\rm sil} = - \dot{\Sigma}_{\rm d}$ by
the conservation of mass.

In the plate-based model, we adopt the same condensation formulation but modify $\dot{\Sigma}_{\rm d}$ (Eq.~\ref{eq:dust-ss-term}) taking into account the limited surface area as follows:
While Eq.~\ref{eq:dust-ss-term} assumes that each particle interacts with the gas phase individually and act as a condensation nucleus, the surface area where particles and gas phase can contact each other may be limited to the surfaces of the solid plate in the plate-based approximation.
We thus modify $\dot{\Sigma}_{\rm d}$, only when calculating the change rate due to condensation, by multiplying $\dot{\Sigma}_{\rm d}$ by the correction factor given by
\begin{equation}
C_{\rm cond} = \min \left(1,\ \frac{S_{\rm plate}}{S_{\rm particle}} = \frac{ 2 \pi r dr}{4 \pi r_{\rm d}^2 N_{\rm d}(r) 2\pi r dr} \right), \label{eq:C-cond}
\end{equation}
where $S_{\rm plate}/S_{\rm particle}$ is the ratio of plate surface area to the summation of surface area for individual particles, and $N_{\rm d}(r)$ is the number density of silicate dust particles.
This diminishes $\dot{\Sigma}_{\rm d}$, inhibiting the vapor from re-condensing immediately outside the silicate line.

\subsection{Equations of Coupled Evolution of Solid-Gaseous Components in the Disc \label{subsec:d-disc-model}}

The governing equations of the surface density of silicate dust particles $\Sigma_{\rm d}$, 
that of silicate vapor $\Sigma_{\rm g}^{\rm sil}$, and the number density of silicate dust $N_{\rm d}$ are 
given respectively by \citep{Desch+2017}
\begin{align}
 &\frac{\partial \Sigma_{\rm d}}{\partial t} + \frac{1}{r} \frac{\partial}{\partial r} \biggl(r\Sigma_{\rm d} \v_{\rm d}
 - r D_{\rm d}\Sigma_{\rm g}^{\rm all}\frac{\partial}{\partial r}  \biggl(\frac{\Sigma_{\rm d}}{\Sigma_{\rm g}^{\rm all}}\biggr)\biggr) 
= \dot{\Sigma}_{\rm d} \label{eq:sigma-d}\\
&\frac{\partial \Sigma_{\rm g}^{\rm sil}}{\partial t} + \frac{1}{r} \frac{\partial}{\partial r} \biggl(r\Sigma_{\rm g}^{\rm sil} \v_{\rm g}
 - r D_{\rm g}(\Sigma_{\rm g}^{\rm all})\frac{\partial}{\partial r}  \biggl(\frac{\Sigma_{\rm g}^{\rm sil}}{\Sigma_{\rm g}^{\rm all}}\biggr)\biggr) 
= - \dot{\Sigma}_{\rm d} \label{eq:sigma-gas1}\\
&\frac{\partial N_{\rm d}}{\partial t} + \frac{1}{r} \frac{\partial}{\partial r} \biggl(rN_{\rm d} \v_{\rm d}
 - r D_{\rm d}N_{\rm g}^{\rm all}\frac{\partial}{\partial r}  \biggl(\frac{N_{\rm d}}{N_{\rm g}^{\rm all}}\biggr)\biggr) \label{eq:N-d}
= 0,
\end{align}
where $\v_{\rm g}$ and $ \v_{\rm d}$ are respectively the radial advection velocities of gas and dust particles, 
$D_{\rm g}$ and $D_{\rm d}$ are their diffusivites, 
which are described in details in Sec.~\ref{subsec:g-advection} to Sec.~\ref{subsec:g-d-diff},
and $\dot{\Sigma}_{\rm d}$ represents the phase change (sublimation/condensation) rate
given by Eq.~(\ref{eq:dust-ss-term}).
We do not include a source/sink term for the equation of $N_{\rm d}$ because we neglect nucleation due to condensation, collisional fragmentation, and coalescence for simplicity.
In the above equations, $\Sigma_{\rm g}^{\rm all} =\Sigma_{\rm g}^{*} + \Sigma_{\rm g}^{\rm sil}$,
where we add a sufficiently small background gas surface density $\Sigma_{\rm g}^{*}$ 
to numerically stabilize single component simulations.

The coupled evolution of $\Sigma_{\rm d}$ and $N_{\rm d}$ enables us to 
 calculate the size evolution of the silicate dust particles due to sublimation and condensation. The particle size strongly affects the drift velocity.
 In our one-dimensional calculation, we adopt the single-size approximation
 \citep{Ormel2014,Sato+2016} where the particle size, $r_{\rm d}$, is represented by a single size,
 depending on the radial distance $r$.

Assuming vertically isothermal structure, 
the mid-plane gas density is given by $\rho_{\rm g} = \Sigma_{\rm g}^{\rm all}/ \sqrt{2 \pi}H_{\rm g}$.

\subsection{Gas Advection \label{subsec:g-advection}}
The gas advection velocity due to viscous diffusion is given by \citep[e.g.,][]{Hartmann2008}
\begin{align}
\v_{\rm g} &= - \frac{3\nu}{r} \frac{\partial \ln(\sqrt{r}\, \nu \Sigma_{\rm g}^{\rm all})}{\partial r} \nonumber \\
&= - \frac{3}{2} \left( 3 - 2 q \right) \ \alpha \
\left( \frac{H_{\rm g}}{r} \right)^2 \v_{\rm K}, \label{eq:v-gas}
\end{align}
where inward gas flow has $\v_{\rm g}<0$ and we use
\begin{align}
q = - \frac{\partial \ln \Sigma_{\rm g}^{\rm all}}{\partial \ln r},
\end{align}
$\v_{\rm K}$ is the Kepler velocity, and $\nu = \alpha H_{\rm g} c_{s} = \alpha (H_{\rm g}/r)^2 \v_{\rm K}r$, which is $\propto r$
in our disc model.
Steady inward accretion, $\dot{M}_{\rm g} = -2\pi r \v_{\rm g} \Sigma_{\rm g}^{\rm all}$ ($\dot{M}>0$ if the flow is inward) with constant $\v_{\rm g} =- 3\nu/2r$, is established for $q=1$.
If $q>3/2$, $\v_{\rm g}>0$ (outward flow). 
Near the silicate line, outward flow is usually obtained.

In general, the vapor surface density distribution inside the sublimation line tends to relax to steady state with $q=1$ and $\dot{M}_{\rm g} = \dot{M}_{\rm PR}$, while the static distribution with $q=3/2$ tends to extend beyond the sublimation line
(for details, see \citet{Okamoto+2022}).
We note that if the positive feed-back is strong enough, the steady accretion 
is not established and the distribution is time dependent.
As we will show later, the re-condensation effect maintains the distribution beyond the sublimation line
steeper than $q=3/2$, even in the case of steady accretion.

\subsection{Radial Drift of Dust Particles\label{subsec:d-drift}}
The drift velocity of a dust particle, $\v_{\rm d}$, is given by 
a sum of the contributions from the PR drag, $\v_{\rm PR}$, and the aerodynamic gas drag, $\v_{\rm GD}$:
\begin{align}
\v_{\rm d} = \v_{\rm GD} + \v_{\rm PR}.
\end{align}

In our particle-based model, $\v_{\rm GD}$ is given by \citep{I&G2016, S&O2017, Hyodo+2019}
\begin{align}
 & \v_{\rm GD} = -\frac{2\mathrm{St} \,\eta \,\v_{\rm K} }{\mathrm{St}^{2} + (1+ Z)^{2}} + \frac{1+Z}{\mathrm{St}^{2} + (1+ Z)^{2}} \v_{\rm g}, \label{eq:v-gas-drag} \\
 & \eta = -\frac{1}{2}\left( \frac{H_{\rm g}}{r} \right)^2 \frac{\partial \ln P_{\rm g}^{\rm all}}{\partial \ln r}, \label{eq:eta}
\end{align}
where $\mathrm{St}$ is the Stokes number of the silicate dust particles and $Z$ is the solid-to-gas density ratio involved in the angular momentum exchange (see below for more exact definition).
The first and second terms in $\v_{\rm GD}$ represent the drift relative to disc gas radial motion
due to angular momentum loss by gas drag
and the drift dragged by the disc gas radial motion, respectively.
Assuming the equation of state for ideal gas, $P_{\rm g}^{\rm all} \propto \rho_{\rm g} T \propto \Sigma_{\rm g}^{\rm all} T/H_{\rm g}$, and $T\propto r^{-1/2}$, 
\begin{align}
\frac{\partial \ln P_{\rm g}^{\rm all}}{\partial \ln r} = -(q+7/4). \label{eq:P_grad}
\end{align}
Using this relation and Eq.~(\ref{eq:v-gas}), Eq.~(\ref{eq:v-gas-drag}) is written as
\begin{align}
& \v_{\rm GD} = -\frac{(H_{\rm g}/r)^2}{\mathrm{St}^{2} + (1+ Z)^{2}}
\left[\left(q+\frac{7}{4}\right)\mathrm{St} + (1+Z) \left(- 3q + \frac{9}{2}\right) \alpha \right] \v_{\rm K}.
\label{eq:vGD}
\end{align}

In the previous treatments in protoplanetary disc studies \citep{Hyodo+2019, Hyodo+2021}, the back-reaction of the dust particle concentration is represented by the midplane dust-to-gas density, $Z=\rho_{\rm d}/\rho_{\rm g}$.
This corresponds to the limit where the angular momentum is exchanged between dust and gas only within the dust particle scale height, $H_{\rm d}$.
The dust particle scale height is given by
\begin{equation}
H_{\rm d}= H_{\rm g} \sqrt{\frac{\alpha}{\mathrm{St} + \alpha}}, \label{eq:Hd}
\end{equation}
where $H_{\rm g}$ is given by Eq.~(\ref{eq:Hgas}), $\alpha$ is 
the turbulent parameter for diffusion.
On the other hand, if the vertical turbulence of gas is no less  effective than viscous diffusion (e.g., $\alpha_{z} \ga \alpha_{\rm acc}$), it transports and mixes the angular momentum from dust particles to the upper gas layer. 
In that case, $Z$ can be rewritten as
\begin{align}
Z = \frac{ \int^{H_{\rm E}}_{-H_{\rm E}} \rho_{\rm d}(z) \,dz} {\int^{H_{\rm E}}_{-H_{\rm E}} \rho_{\rm g}(z) \, dz}, \label{eq:Z}
\end{align} 
where we represent the vertical thickness of the gas layer that can exchange angular momentum with dust particles as $H_{\rm E}$.
Although the exact value of $H_{\rm E}$ is uncertain, we assume in this paper, that velocity shear is restricted within to the Ekman layer of vertical thickness \citep{Metzger+2012},
$H_{\rm E}  \sim  (\nu/\Omega)^{1/2} \sim 0.03 (\alpha/10^{-3})^{1/2} H_{\rm g} $\footnote{In the estimation of the Ekman layer of vertical thickness, \citet{Metzger+2012} adopted the molecular shear viscosity for $\nu$. In that case, Richardson number is so small that Kelvin-Helmholtz (KH) instabilities would occur. We could estimate the vertical thickness of KH-layer so that Richardson number $\sim 1$ using Eq.~(9) in \citet{Metzger+2012} and Eq.~(\ref{eq:Hgas}); $H_{\rm E} \sim H_{\rm g} (H_{\rm g}/r)^{1/2} \sim H_{\rm g} (10^{-3})^{1/2} \sim 0.03 H_{\rm g}$.}.
The vertical thickness of $H_{\rm E}$ determines the efficiency of the angular momentum exchange and significantly affects the behavior of the volatile gas disc through the back-reaction of dust to gas (Sec.~\ref{sec:evo-w-vol}).
Nevertheless, as shown later, our drift velocity formula has similar dependence to that derived by \citet{Metzger+2012} as long as $Z >\mathrm{St}$ and $Z \gg 1$, which is the case in most of our runs and choice of $H_{\rm E}$.

When $H_{\rm E} >H_{\rm d}$, 
Equation~(\ref{eq:Z}) is given by
\begin{align}
Z \simeq \frac{ \Sigma_{\rm d}} {f\Sigma_{\rm g}^{\rm all}}
\end{align}
where
\begin{align}
f \equiv \min \left( 1, \frac{H_{\rm E}}{H_{\rm g}}\right)
\simeq \min \left[ 1, \, 0.03 \left(\frac{\alpha}{10^{-3}} \right) \right].
\label{eq:f}
\end{align}
As we show below, this prescription makes the silicate mass accretion rate onto the star independent of the vertical structure of the solid disc, which may be consistent with \citet{Metzger+2012}.
In our simulation, $\mathrm{St}$ is generally $> 1$, so that  $H_{\rm d} \la H_{\rm E} < H_{\rm g}$ (Eq.~\ref{eq:Hd}).
Note that the settling due to the particle collision and back-reaction of dust to vertical turbulent diffusion is neglected in our model.
However, since the silicate accretion rate is regulated by $H_{\rm E}$ but not by $H_{\rm d}$ in our prescription, this neglect does not affect the results.

Equation~(\ref{eq:vGD}) shows that for $\mathrm{St} > (1+Z) \alpha$, the radial drift relative to gas motion dominates over that associated with the disc gas accretion.
The maximum drift velocity is $\simeq -\eta \v_{\rm K}$ at $\mathrm{St}\sim 1$ and $Z \ll 1$.
For $H_{\rm g}$ given by Eq.~(\ref{eq:Hgas}), it is independent of $r$ because
\begin{align}
-\eta \, \v_{\rm K} & \simeq
-\frac{q+7/4}{2} \left(\frac{H_{\rm g}}{r}\right)^2 \v_{\rm K} \sim -70\ {\rm cm/s},
\label{eq:eta-vk}
\end{align}
where the Kepler velocity $\v_{\rm K}$ is given by
\begin{align}
& \v_{\rm K} \simeq 3 \times 10^{7} \left(\frac{r}{R_\odot}\right)^{-1/2} {\rm cm/s}.
\end{align}

The stokes number $\mathrm{St}$ is given by $\mathrm{St} = t_{\rm s}\Omega$ where the stopping time  $t_{\rm s}$ is the timescale for a particle to lose relative momentum through the gas drag, which is defined by
\begin{equation}
t_{\rm s}=
\left\{
\begin{array}{ll}
\displaystyle \frac{\rho_{\rm int} r_{\rm d}}{\v_{\rm th}\rho_{\rm g}^{\rm all}}  
 & \displaystyle [{\rm Epstein}: r_{\rm d} < \frac{9}{4}\lambda_{\rm mfp}] \\
\displaystyle \frac{4 \rho_{\rm int} r_{\rm d}^2}{9 \v_{\rm th}\rho_{\rm g}^{\rm all} \lambda_{\rm mfp}}  
 & \displaystyle [{\rm Stokes}: \frac{9}{4}\lambda_{\rm mfp}<  r_{\rm d} <  \frac{c_{s}}{\v_{\phi, \rm rel}}]  \\
\displaystyle \frac{8 \rho_{\rm int} r_{\rm d}}{3  \v_{\phi, \rm rel} \rho_{\rm g}^{\rm all}} 
 & \displaystyle [{\rm Newton}: r_{\rm d} > \frac{c_{s}}{\v_{\phi, \rm rel}} \lambda_{\rm mfp}],
\end{array} \label{eq:St}
\right.
\end{equation}
where $\lambda_{\rm mfp}=\mu_{\rm g} /(\sqrt{2} \rho_{\rm g}^{\rm all} \sigma_{\rm mol})$ is the mean free path of the gas, 
$\rho_{\rm int}$ and $\rho_{\rm g}^{\rm all}$ are the particle internal density and the gas spatial density, respectively,
$\v_{\phi, \rm rel}$ is the azimuthal velocity
difference between vapor and dust, equal to $\eta \, \v_{\rm K}$ (eq.~\ref{eq:eta-vk}),
$\sigma_{\rm mol}$ is the molecular collision cross-section, and $\mu_{\rm g}$ is the mean molecular weight of the silicate vapor, which we adopt $\mu_{\rm g} = 30 m_{\rm p}$.
We assume $\sigma_{\rm mol}=5.0 \times 10^{-15}$cm$^{2}$ for silicate vapor, which is a slightly larger than that for molecular hydrogen \citep{Chapman+1970}.
From $R_\odot \sim 7 \times 10^{10}$ cm and Eq.~(\ref{eq:Hgas}),
\begin{align}
\lambda_{\rm mfp} & = \frac{\mu_{\rm g}}{\sqrt{2} \rho_{\rm g}^{\rm all} \sigma_{\rm mol}}
= \frac{\sqrt{\pi} \mu_{\rm g}H_{\rm g}}{\sigma_{\rm mol}\Sigma_{\rm g}^{\rm all}} \\
& \sim 0.5 \left( \frac{\Sigma_{\rm g0}^{\rm all}}{1\  \rm g\ cm^{-2}} \right)^{-1} 
 \left( \frac{r}{r_0}\right)^{q+5/4} \ \rm cm, \label{eq:mfp}
\end{align}
where we set $\Sigma_{\rm g}^{\rm all} = \Sigma_{\rm g0}^{\rm all} \ (r/r_0)^{-q}$ and $r_0 = 0.4 R_\odot$.
In our calculations, $r_{\rm d} \sim \lambda_{\rm mfp}$ and 
dust particles experience the Epstein drag or Stokes drag, depending on $r$ and the evolution of  $\Sigma_{\rm g}^{\rm all}$.
In the former case, substituting $\v_{\rm th}/\v_{\rm K} = \sqrt{8/\pi}\ H_{\rm g}/r$ and
$\Sigma_{\rm g}^{\rm all} = \sqrt{2\pi} \rho_{\rm g}^{\rm all} H_{\rm g}$
into the first equation of Eq.~(\ref{eq:St}), we obtain
\begin{align}
    \mathrm{St} = \frac{\pi}{2} \frac{\rho_{\rm int} r_{\rm d}}{\Sigma_{\rm g}^{\rm all}}.
    \label{eq:St_Sig}
\end{align}
In the Stokes regime, $\mathrm{St}$ is independent of ${\Sigma_{\rm g}^{\rm all}}$.

The mass accretion flux due to the gas drag is given by
\begin{align}
    \dot{M}_{\rm GD} & = - 2\pi r \Sigma_{\rm d} \v_{\rm GD} =
    C_{\rm St-Z} \times 2\pi r \Sigma_{\rm d} \left(\frac{H_{\rm g}}{r}\right)^2 \v_{\rm K} \label{eq:M_GD} 
\end{align}
where $\Sigma_{\rm d,thick}$ is defined by Eq.~(\ref{eq:Sigma_d_thick}), and
\begin{align}
C_{\rm St-Z} = \frac{1}{\mathrm{St}^{2} + (1+ Z)^{2}}
\left[\left(q+\frac{7}{4}\right)\mathrm{St} + (1+Z) \left(- 3q + \frac{9}{2}\right) \alpha \right].
\label{eq:C_St_Z}
\end{align}
For the range where back-reaction is negligible (i.e., $\mathrm{St} \gg Z \rightarrow 0$),  $\dot{M}_{\rm acc}$ scales with $\Sigma_{\rm d}$.
However, for $Z > \mathrm{St}$, the back-reaction alters the dependence. The relation $Z > \mathrm{St}$ holds for almost all solid-enriched regions in our simulations.
In particular, the radial drift relative to gas motion dominates over that associated with the disc gas accretion, for $\mathrm{St} > (1+Z)\alpha$, $C_{\rm \mathrm{St}-Z} \sim (q+7/4) \mathrm{St}/Z^{2}$.
Therefore, Eq.~(\ref{eq:M_GD}) is reduced to
\begin{align}
\dot{M}_{\rm GD}  & \simeq  
C_{\rm GD} \left( q + \frac{7}{4} \right)  \left(\frac{H_{\rm g}}{r}\right)^2 2\pi r\Sigma_{\rm g} \, ; \label{eq:M_GD-BKR} \\
C_{\rm GD} & \equiv \left(\frac{f \Sigma_{\rm g}}{\Sigma_{\rm d}}\right)\ f \, \mathrm{St} \, \v_{\rm K}, 
\label{eq:M_GD-BKR_CDG}
\end{align}
where we used Eq.~(\ref{eq:Hd}) and $Z=\Sigma_{\rm d}/(f \Sigma_{\rm g})$.
Our gas-drag-drift mass accretion rate exhibits the same dependence on 
$\Sigma_{\rm g}$ as Eq.~(\ref{eq:M_GD-M12-comp}) adopted in \citet{Metzger+2012}
\footnote{While  $\mathrm{St}$ is proportional to $\Sigma^{\rm -1}_{\rm g}$ in the Epstein regime, $St$ is independent of $\Sigma_{\rm g}$ in the Stokes regime. Our simulation show that the Stokes regime occurs in the late stage of runaway accretion (Sec. \ref{subsec:M12}), for example.}.
The remaining term $C_{\rm GD}$ is different from the corresponding
term $C_{\rm GD,M12}$ in Eq.~(\ref{eq:M_GD-M12-comp}).
The difference mainly comes from the uncertainty related to the coupling strength between solid and gas-phase in each model, but 
they take close values in the numerical simulations in Sec.~\ref{subsec:M12} and Sec.~\ref{app:M12}.

In the comparative simulation based on the plate-approximation, 
we calculate the gas-drag drift velocity following the drag force formula that \citet{Metzger+2012} derived.
Using their drag force per unit surface area $f_a$, the silicate mass accretion rate due to gas drag is given by \citep{Metzger+2012}\footnote{Note that the definition of $\eta$ in \citet{Metzger+2012} is different from ours;  $\ \eta_{\rm M, 12} = -\frac{1}{2}\frac{\partial \ln P_{\rm g}^{\rm all}}{\partial \ln r}$}
\begin{align}
\dot{M}_{\rm GD, M12} &= \frac{4 \pi r f_{\rm a}}{\Omega} 
=-\frac{\pi r }{\Omega}\ \frac{\partial \ln P_{\rm g}^{\rm all}}{\partial \ln r}  \left|-\frac{\partial \ln P_{\rm g}^{\rm all}}{\partial \ln r} \right|  \frac{ c_{\rm s}^{3}}{Re_{\ast} \Omega r^{2}} \Sigma_{\rm g}\left(1-e^{-\tau_{\rm d}} \right) \label{eq:M_GD-M12}.
\end{align}
For $\tau_{\rm d} \gg 1$, it is reduced to 
\begin{align}
\dot{M}_{\rm GD, M12} & \simeq C_{\rm GD,M12}\left(q+\frac{7}{4} \right)  \left(\frac{H_{\rm g}}{r} \right)^2  2 \pi r \Sigma_{\rm g} \, ; \label{eq:M_GD-M12-comp}\\
C_{\rm GD,M12} & = 
\left| q+\frac{7}{4} \right| \frac{1}{Re} c_{\rm s},
\end{align}
where we have used Eq.~(\ref{eq:Hgas}). 
This is the same as Eq.~(\ref{eq:M_GD-BKR}) in our particle-based model except for $C_{\rm GD}$ and $C_{\rm GD,M12}$. 
The corresponding drift velocity is
\begin{align}
\v_{\rm GD, M12} = - \dot{M}_{\rm GD, M12} /2\pi r \Sigma_{\rm d}. \label{eq:vGD-M12}
\end{align}

We calculate $\v_{\rm PR}$ as
\begin{align}
\v_{\rm PR} = - \dot{M}_{\rm PR}/2\pi r \Sigma_{\rm d}.  \label{eq:vPR}
\end{align}
The silicate mass accretion rate due to Poynting-Robertson effect is given by \citep{Rafikov2011},
\begin{align}
\dot{M}_{\rm PR}=\phi\frac{L_{\rm WD}}{c^{2}}(1-\exp[{-\tau_{\rm d}/\phi}]), \label{eq:dot-M-PR}
\end{align}
where $L_{\rm WD}= \sigma T_{\rm WD}^4 4\pi R_{\rm WD}^2$ is the white dwarf luminosity, $c$ is the speed of light, and $\phi$ is the incidence angle of radiation on the disc surface at distance $r$ from the star 
given by $\phi = (4/3\pi)(R_{\rm WD}/r)$ (Eq.~(\ref{eq:phi})).
The vertical optical depth, $\tau_{\rm d}$, of the particulate disc 
is given by 
\begin{align}
\tau_{\rm d} 
= \frac{\Sigma_{\rm d}}{(4\pi/3) \rho_{\rm int} r_{\rm d}^3}\pi r_{\rm d}^2
= \frac{3\Sigma_{\rm d}}{4 \rho_{\rm int}r_{\rm d}}. \label{eq:tau-d}
\end{align}
The optical depth encountered by incident photons as they traverse the full disc thickness is $\tau_{\rm d}/\phi$. The factor $(1-\exp[{-\tau_{\rm d}/\phi}])$ in Eq.~(\ref{eq:dot-M-PR}) characterizes the efficiency of radiation absorption by the disc.
The dust particle surface density for optically thick condition $\tau_{\rm d}/\phi \ga 1$
is given by
\begin{align}
\Sigma_{\rm d} & \ga \Sigma_{\rm d,thick} \equiv \frac{16}{9\pi} \rho_{\rm int} r_{\rm d} \left(\frac{r}{R_{\rm WD}}\right)^{-1} \\
 & \simeq 4.2 \times 10^{-2} \left(\frac{r_{\rm d}}{1 \rm cm}\right) \left( \frac{R_{\rm WD}}{0.01 R_\odot} \right) \left( \frac{r}{r_0} \right)^{-1} \, {\rm g/cm^2} \label{eq:Sigma_d_thick}.
\end{align}

In the optically thin limit ($\tau_{\rm d}/\phi \ll 1; \Sigma_{\rm d} \ll \Sigma_{\rm d,thick}$), 
\begin{align}
\dot{M}_{\rm PR,thin} & \simeq \frac{\tau_{\rm d} L_{\rm WD}}{c^{2}} \\
 & \simeq 4 \times 10^{7} \left(\frac{\Sigma_{\rm d}}{\Sigma_{\rm d,thick}}\right) \left(\frac{r_{\rm d}}{1 \rm cm}\right)^{-1}
\biggl(\frac{L_{\rm WD}}{L_{\rm WD,*}} \biggr)\left( \frac{r}{r_0} \right)^{-1} \ \rm g/s.
\label{eq:Mdot_PR_thin}
\end{align}
where we scale $L_{\rm WD}$ by $L_{\rm WD,*}$, the value of $L_{\rm WD}$ for $T_{\rm WD} = 10^4\ {\rm K} $ and  
$R_{\rm WD}=0.01 R_{\odot}$.
The corresponding drift velocity is
\begin{align}
\v_{\rm PR,thin} & = -\frac{\dot{M}_{\rm PR}}{2\pi r \Sigma_{\rm d}}=-\frac{3}{8\pi} \frac{L_{\rm WD}}{\rho_{\rm int} r_{\rm d} c^2 r} \\
           & \simeq 6 \times 10^{-3} \ \biggl(\frac{r_{\rm d}}{0.01 {\rm cm}}\biggr)^{-1} \ \biggl(\frac{L_{\rm WD}}{L_{\rm WD,*}} \biggr)
           \left( \frac{r}{r_0} \right)^{-1} \ {\rm cm/s}, \nonumber
\end{align}
This corresponds to the PR drift velocity of the fully exposed particles \citep{Burns+1979}.
In the optically thick case ($\tau_{\rm d}/\phi \ga 1; \Sigma_{\rm d} \ga \Sigma_{\rm d,thick}$), 
Eq.~(\ref{eq:dot-M-PR}) is reduced to
\begin{align}
\dot{M}_{\rm PR,thick} &= \phi\frac{L_{\rm WD}}{c^{2}}  \nonumber \\
                        &\simeq 4 \times 10^{7}\ 
                        \biggl(\frac{L_{\rm WD}}{L_{\rm WD,*}} \biggr) \biggl(\frac{R_{\rm WD}}{0.01 R_\odot} \biggr) \biggl(\frac{r}{r_0} \biggr)^{-1} \ {\rm g/s}.
                        \label{eq:PR-limit}
\end{align}
Because $\dot{M}_{\rm PR,thick}$ is independent of $\Sigma_{\rm d}$,
$\dot{M}_{\rm PR}$ does not increase any more with $\Sigma_{\rm d}$ once the disc
becomes optically thick, so that
$\dot{M}_{\rm PR,thick}$ is the upper limit of the accretion rate
in the case of the PR drag alone. 

When $\dot{M}_{\rm PR}$ is considered as the silicate vapor source term, 
it is evaluated by $\dot{M}_{\rm PR,thick}$ at the silicate line radius
if the region near the silicate line is optically thick.
Otherwise, it is evaluated by $\dot{M}_{\rm PR,thick}$ at the 
inner boundary of optically thick region.

\subsection{Diffusion \label{subsec:g-d-diff}}

 In the basic equation for solid particle evolution given by Eq.~(\ref{eq:sigma-d}),
the particle radial diffusion is included,
while the particle diffusion in the vapor disc was not considered
in the previous studies for the WD disc evolution modeling \citep{Bochkarev+2011, Metzger+2012}.
For consistency, we introduce the particle diffusion.

We identify diffusivity of silicate vapor as the turbulent viscosity,
\begin{align}
D_{\rm g} = \alpha c_{\rm s} H_{\rm g} \simeq \alpha \left(\frac{H_{\rm g}}{r}\right)^2 \v_{\rm K} r.  \label{eq:Dg}
\end{align}
The solid particle diffusion by drag from turbulent gas is given by
\citep{Youdin+2007}
\begin{align}
D_{\rm d} = \frac{D_{\rm g}}{1+\mathrm{St}^{2}}.
\label{eq:Dg0}
\end{align}
\citet{Hyodo+2019} and \citet{Ida+2021} discussed the back-reaction to the particle
diffusivity (the $Z$-dependence of $D_{\rm d}$).
Although the dependence is not clear, they proposed that the correction factor
for the back reaction is $\propto (1+Z)^{-1}$ in the case of $\mathrm{St} \ll 1$.
Taking into account both the effect of $\mathrm{St}$ and the back-reaction
and following the form of exact derivations of the advection terms (Eq.~(\ref{eq:vGD})),
we adopt
\begin{align}
D_{\rm d} & = \frac{1+Z}{\mathrm{St}^{2} + (1+Z)^2} D_{\rm g} \\
 & \simeq \frac{(H_{\rm g}/r)^2}{\mathrm{St}^{2} + (1+Z)^2} (1+Z) \ \alpha \v_{\rm K} r. \label{eq:Dg1}
\end{align}
As we pointed out, because $Z$ can be a few or more orders of magnitude larger than unity,
the back reaction is very important for radial diffusion while vertical diffusion, which determines $H_{\rm d}$, does not affect the results.
Comparison of $D_{\rm d}/r$ with $\v_{\rm GD}$ (Eqs.~(\ref{eq:sigma-d}) and (\ref{eq:vGD})) 
suggests that the diffusion effect may be comparable to the effect of the disc accretion in $\v_{\rm GD}$.


\subsection{Simulation Settings for One-component Silicate Disc \label{subsec:setting}}
We numerically integrate Eq.~(\ref{eq:sigma-d})--(\ref{eq:N-d}) using the finite volume method.
In the disc environment with high temperature and high vapor pressure around white dwarfs, the timescale for condensation $t_{\rm cond}$ can be extremely short compared to that for other processes such as advection and diffusion.
From eq.~(\ref{eq:dust-ss-term}), we can estimate $t_{\rm cond}\sim 1/  (C_{\rm cond}) R_{\rm con} \Sigma_{\rm d}$, and the smallest $t_{\rm cond}$ is a few seconds near the silicate sublimation line.
To avoid such extremely short timesteps, we combine the analytic timestep scheme developed by Matsuura et al.~(in prep.) based on \citet{Clark1973} for the calculation of source/sink term due to sublimation and condensation.
When using the analytic timestep scheme, we have confirmed that the total mass of gas and dust particles is conserved.
We adopt logarithmic grids that extend from the WD surface ($r=R_{\rm WD}\sim 0.01 \ R_\odot$) out to a radius beyond the Roche limit radius ($r=1.1 r_{\rm R} \sim 1.4 \ R_\odot$). The number of radial grids is 100--300, depending on calculation settings.

For the comparison with \citet{Metzger+2012}, we adopt 
the ring-like distribution 
of $\Sigma_{\rm d}$ as an initial condition, which is similar to that used by \citet{Metzger+2012},
\begin{equation}
\Sigma_{\rm d} = \Sigma_{\rm d0} \exp\biggl[-\left(\frac{r-r_{0}}{\Delta r}\right)^{2}\biggr],
\label{eq:initial-sigma-d}
\end{equation}
where $r_{0}$ and $\Delta r$ specify the ring center and its radial width, respectively. In this study, we fix $r_{0} \simeq 0.4  R_{\odot}$ and $\Delta r \simeq 0.04 R_{\odot}$ for all simulations, 
so that the initial ring is located outside the sublimation line at $\simeq 0.1 R_{\odot}$. 
We confirmed that the results hardly change for larger $r_{0}$. 
We choose a value of $\Sigma_{\rm d0}$ for the particle disc to be optically thick  
($> \Sigma_{\rm d,thick}$; Eq.~(\ref{eq:Sigma_d_thick}))
(i.e., $\tau_{\rm d0}/\phi >1$) 
for all cases. In Sec.~\ref{sec:res-1comp}, we fix the value at $\Sigma_{\rm d0} = 10$ g/cm$^{2}$, and this corresponds to the particle disc mass of $M_{\rm d0} \sim 10^{22}$ g. The value is typical for observationally inferred rocky mass accreted onto WDs \citep{Farihi+2010}.
In Sec.~\ref{sec:evo-w-vol}, we vary $\Sigma_{\rm d0}$ to study the dependence of accretion rate on the silicate to volatile vapor mass ratio for the volatile-rich disc.
%

For the primary purpose of investigating the disc evolution,
we simply assume that all particles initially have $r_{\rm d}=0.1$ or 1 cm for most of the calculations in this study. 
\citet{Graham1990} observationally suggested that most of the solid mass may be contained in cm-sized particles, and these small particles would be generated by collision cascades of fragments of the tidally disrupted planetary bodies \citep[e.g.,][]{Kenyon+2017}.
We take $\alpha=10^{-3}$ for the fiducial value in this paper, although the appropriate value of $\alpha$ is not clear. 
With this value of $\alpha$, we take $f \ge 0.03$ in $H_{\rm E}$ (see Sec.~\ref{subsec:d-drift}).

\section{Evolution of one-component silicate disc} \label{sec:res-1comp}
In this section, we clarify how condensation affects the disc evolution 
and the resultant silicate accretion rate onto the stellar surface 
($\dot{M}_{\rm acc} \equiv \dot{M}_{\rm g}^{\rm sil} |_{r=R_{\rm WD}}$). 
To show that results are essentially independent of gas-drag treatment, we perform the simulation using gas-drag-drift formulas derived from the particle-based approach (Eq.~\ref{eq:v-gas-drag}) and the plate-based approach (Eq.~\ref{eq:vGD-M12}), respectively. 
We first show that, without condensation, the gas-drag drift causes the runaway accretion enhancement \citep{Metzger+2012}, even for the particle-based prescription in Sec.~\ref{subsec:M12}.
Then, in Sec.~\ref{subsec:with-con}, with condensation in each of the models adopting particle-based and plate-based drag formulas, we demonstrate that condensation drastically changes the disc evolution and inhibits runaway accretion in both simulations.
This indicates that the observed range of the accretion rates could not be covered by the accretion from one-component silicate discs.

\subsection{Without Condensation: Runaway Accretion \label{subsec:M12}}
\begin{figure}
\hspace{-2.5mm}
\includegraphics[bb=0 0 677 476, width=\columnwidth]{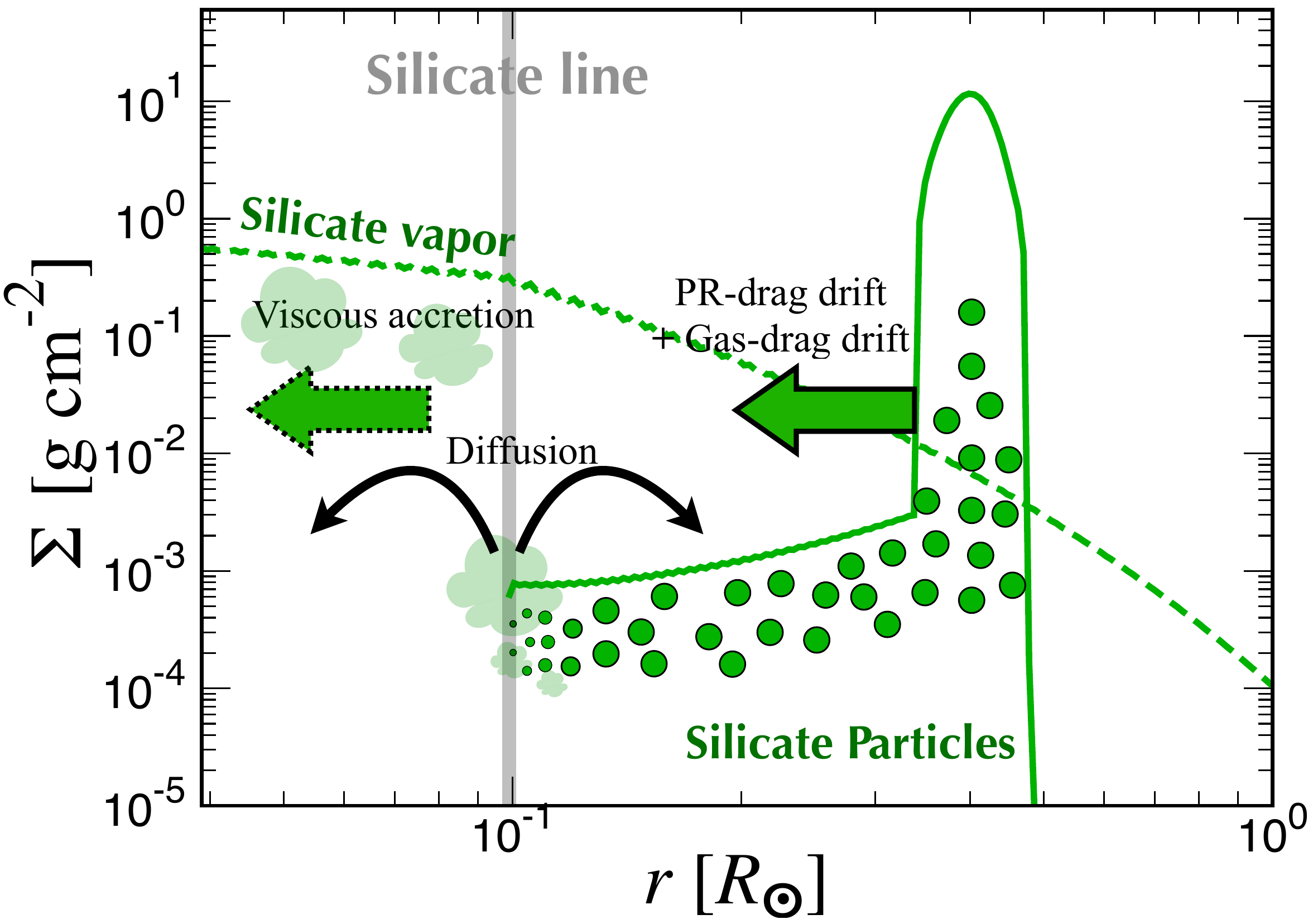}
\vspace{-1mm}
\caption{
The schematic illustration of the snapshot of the solid ($\Sigma_{\rm d}$; solid line) and the vapor ($\Sigma_{\rm g}^{\rm sil}$; dashed line) disc surface density at $t= 10000$ year.  The vertical gray line shows the location of the silicate sublimation line. The horizontal axis is scaled by the Solar radius $R_{\odot}$. 
\label{fig:demo-sigma}}
\end{figure}

To verify that the runaway accretion occurs in our particle-based prescription, 
we perform disc evolution simulations using our model in Sec.\ref{sec:model} with i) turning off the condensation and ii) using the gas-drag-drift formula of Eq.~(\ref{eq:v-gas-drag}). We adopt $r_{\rm d} =0.1$ cm and $f=0.1$.
Figure \ref{fig:demo-sigma} schematically shows the surface density evolution of solid and vapor until $t = 6000$ years.
As dust particles approach the silicate sublimation line, silicate vapor is produced. 
Inside the sublimation line, only the vapor disc is present, because the sublimation timescale is
much shorter than the particle drift timescale.
The produced vapor diffuses inward to establish the steady accretion with $\Sigma_{\rm g} \propto r^{-1}$
and also diffuses outward to expand with $\Sigma_{\rm g} \propto r^{-3/2}$ in an inside-out manner (diffusion timescale is $\propto r$).

\begin{figure}
\hspace{-2.5mm}
\includegraphics[bb = 0 0 360 252, width=\columnwidth]{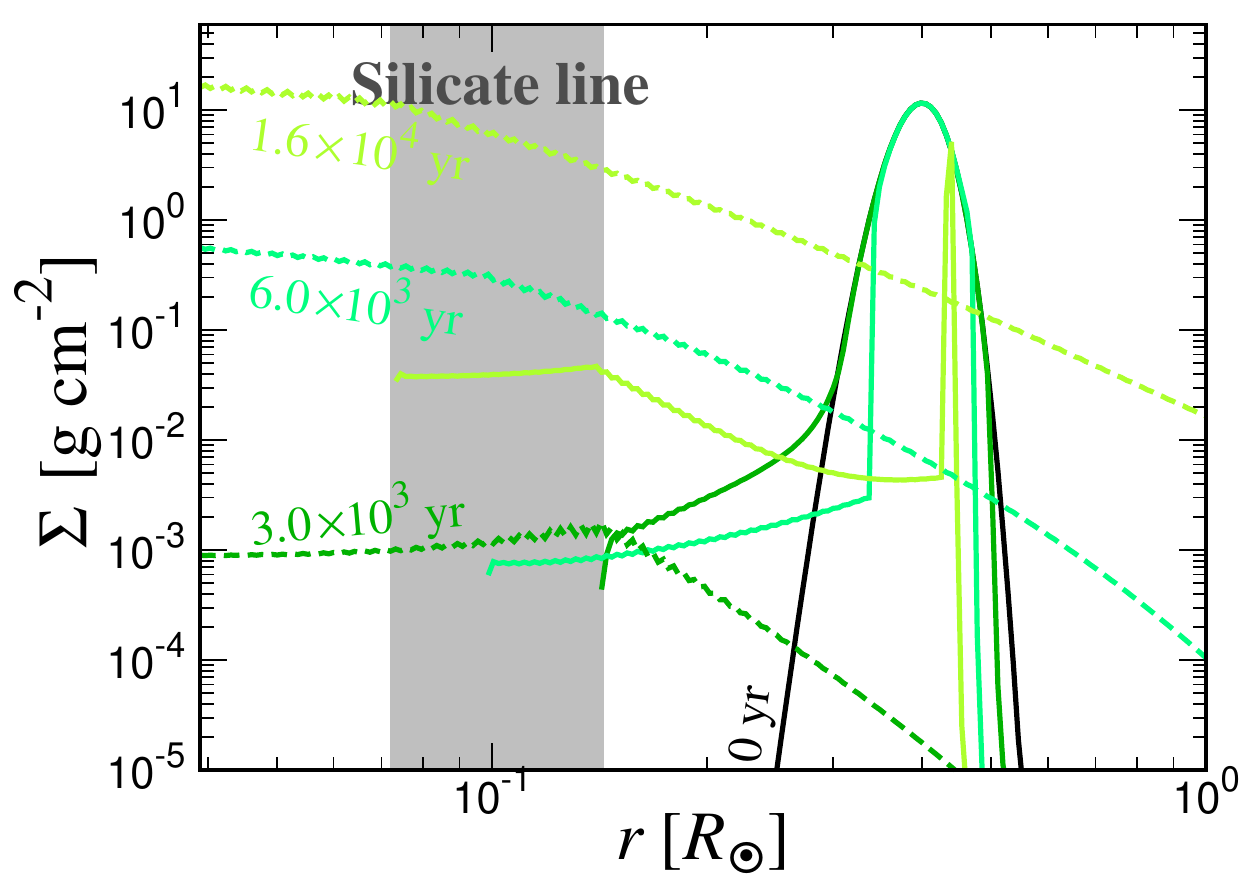}
\vspace{-1mm}
\caption{The surface density evolution of the solid disc ($\Sigma_{\rm d}$; solid lines) and the vapor disc ($\Sigma_{\rm g}^{\rm sil}$; dashed lines) calculated from our particle-based model with turning off the condensation.
We adopted $\alpha=10^{-3}$, $f=0.1$, and $r_{\rm d}=0.1$ cm. The different colors represent the surface density profiles at different times. The sublimation line evolves with time, and hence the gray shaded region represents the area across which the sublimation line passes.  
}
\label{fig:sigma-wo-cond}
\end{figure}

However, this is a quasi-steady/static state and $\Sigma_{\rm g}$ secularly increases with time as a whole, as shown in Fig.~\ref{fig:sigma-wo-cond}.
Outside the silicate line, solid and vapor discs co-exist, and especially at $t > 3 \times 10^3$ years, the inward drift of the silicate particles is accelerated due to increasing $\Sigma_{\rm g}$. 
As we already pointed out, this acceleration is common to plate drag regime (Eq.~\ref{eq:M_GD-M12}) and indispensable for the runaway accretion.
Since the back-reaction is so effective that $Z > \mathrm{St}$ and the gas-drag-drift governs the silicate dust accretion (i.e., $Z < \mathrm{St}/\alpha$) at the inner edge of optically thick region ($r \sim 0.3 R_{\odot}$), the silicate accretion flux follows Eq.~(\ref{eq:M_GD-BKR}) to increase with the increase of  $\Sigma_{\rm g}$.
Once the $\Sigma_{\rm g}^{\rm sil}$ exceeds a critical value such that $\v_{\rm GD} > \v_{\rm PR}$, all solid particles rapidly drift into the sublimation line.

\begin{figure}
\hspace{-2.5mm}
\includegraphics[bb= 0 0 360 252, width=\columnwidth]{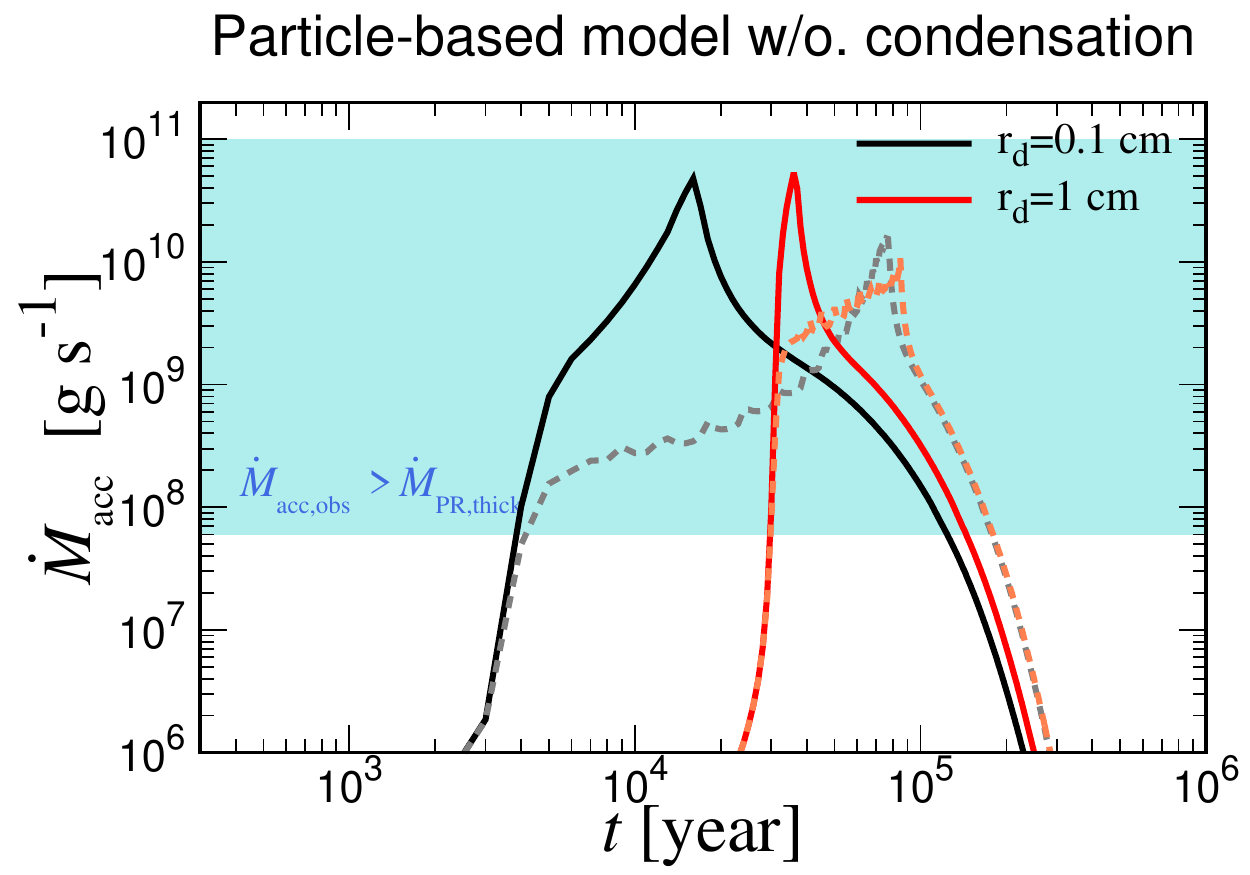}
\vspace{-1mm}
\caption{
Time evolution of the accretion rate $\dot{M}_{\rm acc} \equiv \dot{M}_{\rm g}^{\rm sil}|_{r=R_{\rm WD}}$ calculated from our particle-based model with turning off the condensation.
The solid and dotted lines show results with $f = 0.1$ cm and $f = 0.03$.
The cyan region highlight the observed values $\dot{M}_{\rm acc,obs}$ higher than $\dot{M}_{\rm PR,thick}$ evaluated at the inner boundary of optically thick region
(Eq.~(\ref{eq:PR-limit})).
\label{fig:acc_rate-wo-cond}}
\end{figure}

The positive feedback between the drag force and $\Sigma_{\rm g}$ secularly increases $\dot{M}_{\rm acc}$ with time (Fig.~\ref{fig:acc_rate-wo-cond}), resulting in the runaway accretion at $t \gtrsim 2 \times 10^4$ years.
At this time, the gas drag is so efficient that the sublimation supplies gas near the silicate line faster than viscous diffusion to remove the gas \citep{Rafikov2011b,Metzger+2012}. 
Figure \ref{fig:acc_rate-wo-cond} also shows the evolution of accretion rate with various $r_{\rm d}$ and $f$. 
Although the onset timing of runaway accretion and the presence/absence of plateau depend on the parameter choice, all of the peak values of $\dot{M}_{\rm acc}$ are orders of magnitude higher than the rate that can be achieved by PR drag alone.

In Appendix \ref{app:M12}, we perform the same setting simulation as \citet{Metzger+2012} by using the gas-drag-drift formula of Eq.~(\ref{eq:vGD-M12})
to reproduce their results of the surface density evolution (Fig.~\ref{fig:sigma-wo-cond-plate}) and the accretion rate evolution (Fig.~\ref{fig:acc_rate-wo-cond-plate}).
The runaway accretion process shown in these results is qualitatively reproduced by our simulations with the particle drag prescription.
The dust surface density distributions are different between our results and the reproduction results of \citet{Metzger+2012} due to the different gas-drag drift velocity of dust particles, but the overall evolution trend is common.
In the next section, by simulations with condensation, we will show
that the effect of re-condensation of silicate vapor totally suppresses the runway accretion in the case of silicate one-component discs.


\subsection{Importance of Condensation \label{subsec:with-con}}

We present the disc evolution and the accretion rates 
when additionally including condensation into particle-based disc model and \citet{Metzger+2012}'s plate-based disc model, respectively.
In both cases, the suppression of the runaway accretion is caused by condensation.

\begin{figure*}
\includegraphics[bb= 0 0 576 720, scale=0.8]{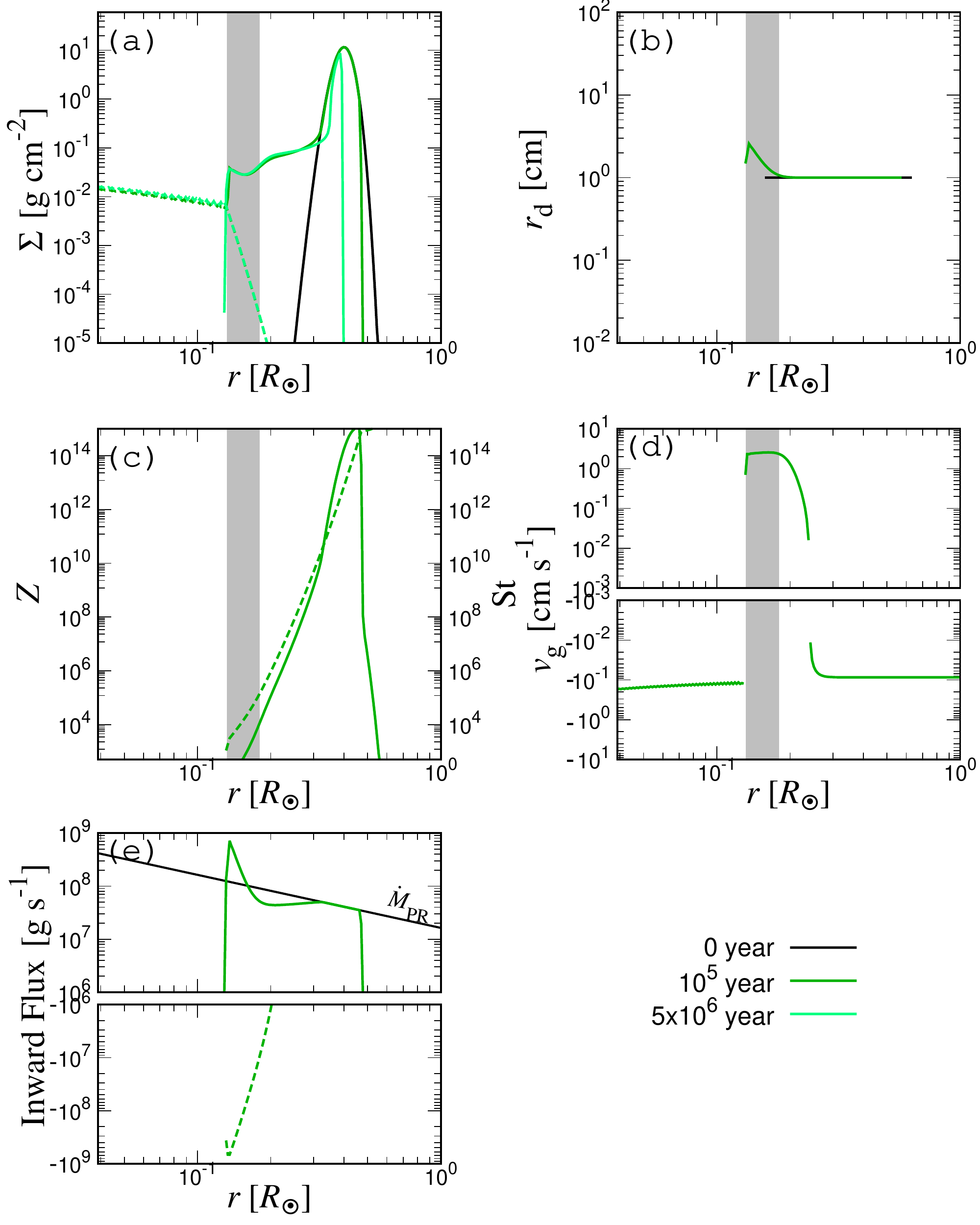}
\caption{
Time evolution of (a) silicate particle surface density $\Sigma_{\rm d}$ (solid lines) and silicate vapor surface density $\Sigma_{\rm g}^{\rm sil}$ (dashed lines), (b) particle radius $r_{\rm d}$, (c) solid concentration $Z$ (solid lines) and $\mathrm{St}$ (dashed lines), (d) gas velocity $\v_{\rm g}$, and (e) inward flux of silicate particles (solid line) and vapor (dashed line) obtained from our particle-based model calculation including condensation. The different colors represent physical quantities at different times. 
We adopt  $\alpha = 10^{-3}, f = 0.1$, and $r_{\rm d} = 1$ cm.
The sublimation line evolves with time, and the gray shaded region represents the area across which the sublimation line passes. In panel (e), the black line shows analytically estimated $\dot{M}_{\rm PR,thick}$ (Eq.~\ref{eq:PR-limit}).
\label{fig:all-w-cond}}
\end{figure*}

Using the drag force calculated from the summation of drag forces for individual particles with back-reaction through the dependence on $Z$ (Eq.~\ref{eq:v-gas-drag}), we perform the disc evolution simulation with turning on the condensation.
Figure~\ref{fig:all-w-cond} shows physical quantities which characterize disc structures in our fiducial simulation.
Compared with the result with turning off the condensation (Fig.~\ref{fig:sigma-wo-cond}),
$\Sigma_{\rm g}^{\rm sil}$ is much smaller both inside and outside the silicate line (Fig.~\ref{fig:all-w-cond}a).
Outside the silicate line, $\Sigma_{\rm g}^{\rm sil}$ (the silicate vapor surface density) decreases with $r$ much rapidly.
As we already pointed out, the re-condensation timescale is significantly shorter than
the radial diffusion timescale, so that $\Sigma_{\rm g}^{\rm sil}$ outside the silicate line is almost completely  
determined by the saturation pressure ($P_{\rm eq}$) given by Eq.~(\ref{eq:P-sat}).
The smaller $\Sigma_{\rm g}^{\rm sil}$ also results in 
$Z > 10^4$ (solid dominated)\footnote{In particular, the optically thick region  ($r\gtrsim0.3 R_{\odot}$) obtains $Z>St$, where the solid accretion rate driven by gas drag follows Eq.~(\ref{eq:M_GD-BKR}).} and $\mathrm{St} > 10^4$ (weak coupling) as shown in Fig.~\ref{fig:all-w-cond}c.
The high $Z$ and $\mathrm{St}$ imply 
lower $\v_{\rm GD}$ due to the gas drag from the vapor (Eq.~\ref{eq:v-gas-drag}).
The solid particle mass flux provided from the Gaussian ring is always limited by that due to the optically-thick limit of PR drag (Fig.~\ref{fig:all-w-cond}e), resulting in one to two orders of magnitude lower $\Sigma_{\rm g}^{\rm sil}$ inside the silicate line.
Therefore, runaway accretion never occurs.

The re-condensation of silicate vapor generates a recycling flow structure at the silicate line, which never appears in the simulations with tuning off condensation, as follows:
Because the saturation vapor pressure produces a sharp negative gradient in $\Sigma_{\rm g}^{\rm sil}$ outside the sublimation line, $\v_{\rm g}$ takes positive values (outward flow) locally as shown in Fig.~\ref{fig:all-w-cond}d.
From Eqs.~(\ref{eq:P-sat}) with $T \sim 2000 \ (r/r_{\rm sil})^{-1/2}\ \rm K$ and Eq.~(\ref{eq:P_grad}),
\begin{align}
q & = - \frac{\partial \ln \Sigma_{\rm g}^{\rm sil}}{\partial \ln r} \simeq 
- r \frac{\partial \ln P_{\rm eq}}{\partial r} - \frac{7}{4} \nonumber \\
 & \sim \frac{\mathcal{A}}{2 \times 2000}\left(\frac{r}{r_{\rm sil}}\right)^{1/2} - \frac{7}{4}
 \sim 14.5,
 \end{align}
and substituting $q\simeq 14.5$ into Eq.~(\ref{eq:v-gas}), we estimate $\v_{\rm g} \sim 2$ cm/s, which explains 
the magnitude of the outward flow in Fig.~\ref{fig:all-w-cond}d.
The outwardly diffused vapor immediately re-condenses if the partial pressure exceeds the saturation pressure.
This grows silicate particle size in the vicinity of the silicate line (Fig.~\ref{fig:all-w-cond}b). 
However, the re-condensed particles drift back to the silicate line due to the gas drag from the silicate vapor in addition to PR-drag, which provides the solid particle mass flux in excess of $\dot{M}_{\rm PR, thick}$ in the vicinity of the silicate line.
As a result, the accretion flux of silicate vapor is equal to the mass flux of solid particles flowing into the silicate line (Fig.~\ref{fig:all-w-cond}e).

\begin{figure}
\hspace{-2.5mm}
\includegraphics[bb = 0 0 360 252, width=\columnwidth]{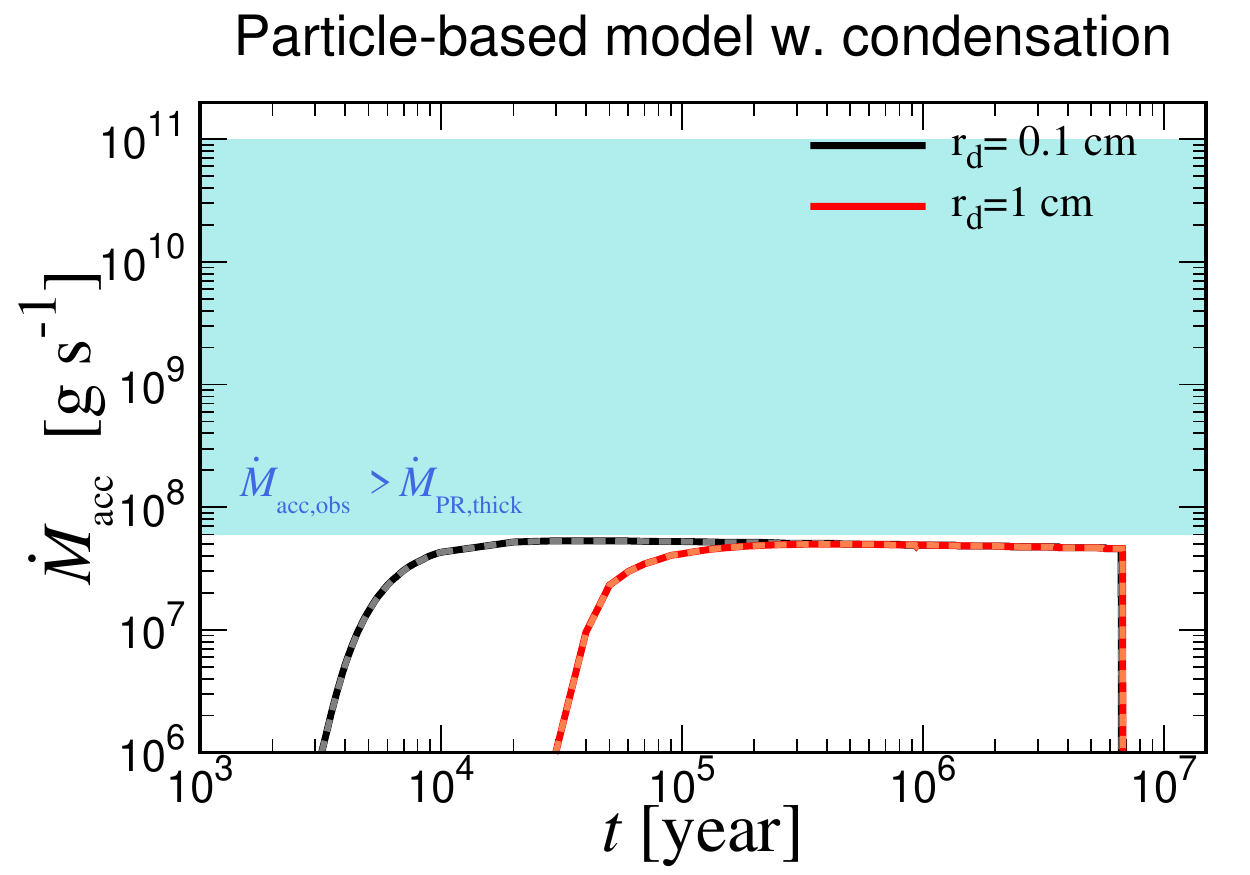}
\vspace{-1mm}
\caption{
Same as Figure \ref{fig:acc_rate-wo-cond} except for turning on the condensation.
Note that solid lines ($f=0.1$) and dashed lines ($f=0.03$) exactly over lapped with each other for same $r_{\rm d}$. 
 }
\label{fig:acc_rate-w-cond}
\end{figure}

Once a steady state without runaway accretion is established, the accretion rate onto the WD is simply equal to that caused by the PR drag alone, $\dot{M}_{\rm PR,thick}$ (Figure.~\ref{fig:acc_rate-w-cond}).
As almost all solid mass accretes onto the star through steady accretion, it continues until $M_{\rm d0}/\dot{M}_{\rm PR,thick} \sim 7 \times 10^6$ year.
In the final stage of accretion, the particle disc becomes optically thin (Eq.~\ref{eq:dot-M-PR}) as the particle size increases due to repeated sublimation and re-condensation, and the PR-drag slows down in proportion to $\tau_{\rm d}$.

The above arguments to explain the accretion rate obtained by our full disc model
is almost independent of the disc parameters, $r_{\rm d}$ and $f$.  
Figure.~\ref{fig:acc_rate-w-cond} shows the results with various $r_{\rm d}$ and $f$
obtained by our particle-based disc model including condensation.
We stress that, in our full disc model, the condensation suppresses the runaway accretion even with the parameters which cause intense runaway accretion in the simulation by turning off the condensation (e.g., $r_{\rm d}$= 1 cm and $f=0.1$ in Fig.~\ref{fig:acc_rate-wo-cond}).
The accretion rate converges to $\dot{M}_{\rm PR, thick}$ for all parameters. 
In addition, as $\dot{M}_{\rm PR, thick}$ is independent of $\Sigma_{\rm d}$ (Eq.~\ref{eq:PR-limit}), $\dot{M}_{\rm acc}$ does not increase even as the increase of silicate disc mass.

We also perform the simulation by adding the condensation effects to the plate-based model that \citet{Metzger+2012} adopted.
We use the gas-drag drift velocity derived from the gas-drag force between the solid plate surface and the layer of gas (Eq~.\ref{eq:vGD-M12}). Furthermore, we modify the surface density change rate due to condensation by multiplying the correction factor, $C_{\rm cond}\le 1$, in Eq.~(\ref{eq:C-cond}).

\begin{figure}
\hspace{-2.5mm}
\includegraphics[bb = 0 0 360 252, width=\columnwidth]{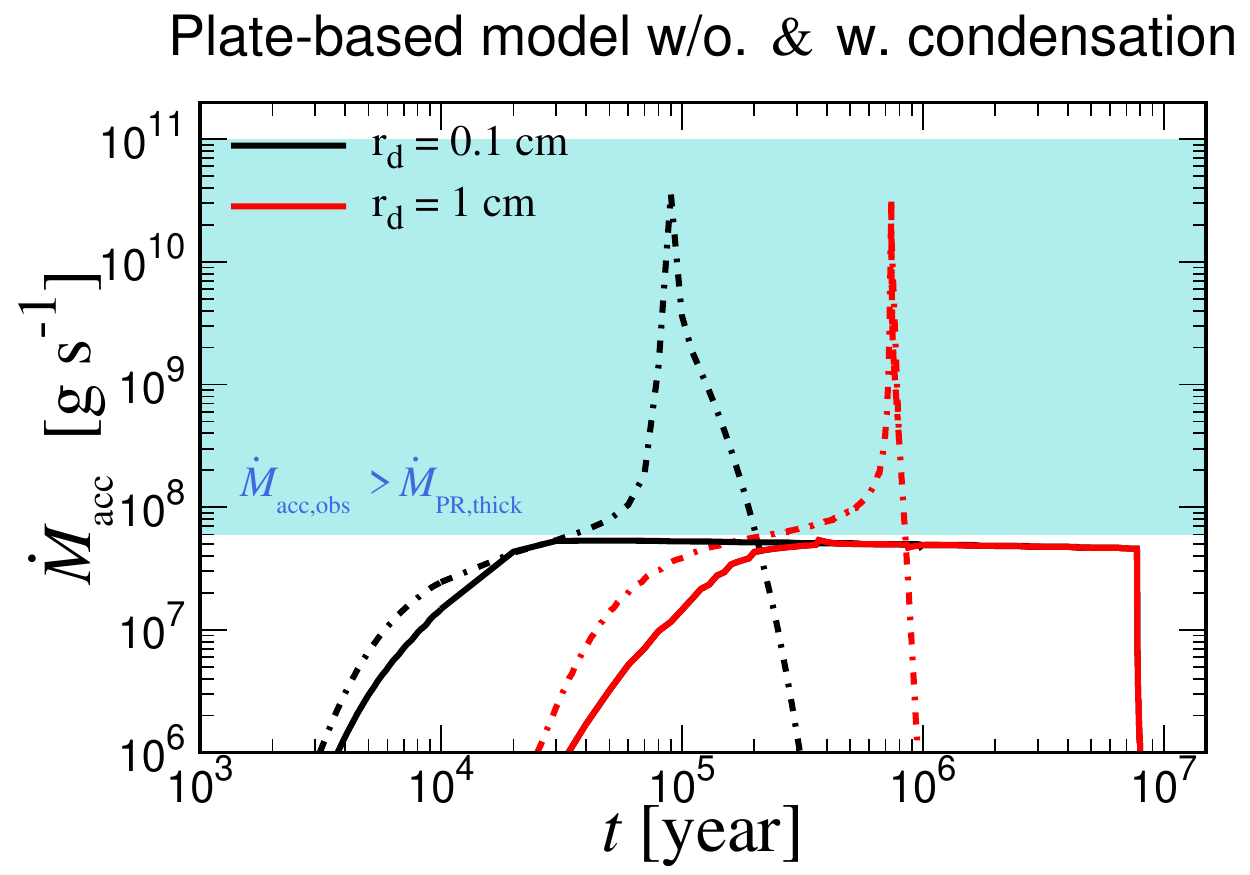}
\vspace{-1mm}
\caption{
Time evolution of the accretion rates $\dot{M}_{\rm acc}$ calculated from plate-prescription model with turning off the condensation (dotted-dash lines) and on the condensation (solid lines).
We adopt the turbulent parameter $\alpha=10^{-3}$ and $Re_{\ast}=0.3$.  
}
\label{fig:acc_rate-w-cond-plate}
\end{figure}

Figure~\ref{fig:acc_rate-w-cond-plate} shows the accretion rate with various $r_{\rm d}$ obtained by the plate-based disc model turning on and off the condensation, respectively. 
We choose $Re_{\ast}=0.3$ so that the sufficiently strong gas-solid feedback can cause the runaway accretion for all parameters (for details, see \citep{Metzger+2012}).
Nevertheless, we find that the accretion rate in the runs with turning on the condensation is limited to $\dot{M}_{\rm PR}$ for all parameters. 
Even if we adopt $\v_{\rm GD}$ derived by the drag force formula in the solid plate approximation (Eq.~\ref{eq:vGD-M12}), the condensation completely halts the runaway accretion.
In addition, the correction of condensation surface area gives a negligible impact on the results because the time scale of condensation is still several orders of magnitude shorter than those of vapor diffusion.
%
%

\section{Evolution of Silicate Discs with Volatile Vapor} \label{sec:evo-w-vol}
In Sec.~\ref{sec:res-1comp}, our results demonstrate that mono-compositional silicate discs cannot produce the high accretion rates observed for WDs with discs due to the effects of re-condensation of silicate vapor outside the silicate line, regardless of disc parameters. 
Even for large mass discs, $\dot{M}_{\rm acc}$ cannot exceed the optically-thick asymptotic value $\sim 10^8$~g/s.


In this section, we investigate the effects of the volatile vapor such as water vapor that does not undergo condensation in the region where silicate particles are distributed.
Based on the dynamical simulations \citep{Mustill+2018, Li+2022}, ice-bearing bodies could infall to add the volatile vapor to the silicate disc.
In particular, we focus on (1) the possibility to provide a higher range
of observed accretion rates due to gas drag from volatile vapor.
We also study (2) whether such a high accretion rate can reproduce rock-rich photospheric composition or not.

In the results part, we present both the evolution of instantaneous accretion rate and accumulated accretion mass.
If the metal-sinking timescale, $t_{\rm sink}$, in the WD atmosphere is significantly short, the photospheric composition would correspond to the silicate-to-volatile mass ratio in accreting flux.
On the other hand, for the longer $t_{\rm sink}$, the time-integrated accretion mass would be a representative of the photospheric composition.
The former would apply to WDs with hydrogen-dominated atmospheres (DA-type), which typically have $t_{\rm sink}$ of 1 day to $10^4$ year \citep{Paquette1986, Koester2009}.
The latter would be the case for WDs with helium-dominated atmospheres (DB-type), which ha $t_{\rm sink}$of $10^{4}-10^6$ year \citep{Paquette1986, Koester2009}.

In Section \ref{subsec:vol-model}, we describe our model of volatile vapor disc evolution, the update part from the previous one in Sec.\ref{sec:model}, and our numerical settings. 
In Sec.~\ref{subsec:one-to-one}, we present the disc evolution and accretion rate of silicate/volatile discs for the initial silicate-to-volatile mass ratio, $M_{\rm disc}^{\rm sil}/M_{\rm disc}^{\rm vol} =1$.
While the disc with $M_{\rm disc}^{\rm sil}/M_{\rm disc}^{\rm vol} =1$ results in ice-rich accretion in the late evolution stage, we demonstrate that discs with $M_{\rm disc}^{\rm sil}/M_{\rm disc}^{\rm vol} =10$ can reproduce silicate-rich composition over the entire evolution period in Sec.~\ref{subsec:ten-to-one}.
Both of $M_{\rm disc}^{\rm sil}/M_{\rm disc}^{\rm vol} =1$ and 10 cases achieve the observed high accretion rate. 

 \subsection{Two-component Disc Model \label{subsec:vol-model}}

 \subsubsection{Equations of volatile vapor evolution in the disc}
We self-consistently introduce the evolution of volatile vapor into our silicate solid/vapor disc model developed in Sec.~\ref{sec:model}.
For the basic disc structures and the evolution of silicate solid/vapor discs, we adopt the same model in Sec.~\ref{sec:model}.
For all of the calculations in Sec.~\ref{sec:evo-w-vol}, we adopt the particle-based approach that enables us to treat the silicate sublimation/condensation calculation self-consistently in the advection/diffusion disc model.

The governing equation of the surface density of volatile vapor, $\Sigma_{\rm g}^{\rm vol}$, is given by \citep{Desch+2017}
\begin{align}
\frac{\partial \Sigma_{\rm g}^{\rm vol}}{\partial t} + \frac{1}{r} \frac{\partial}{\partial r} \biggl(r\Sigma_{\rm g}^{\rm vol} \v_{\rm g}
 - r D_{\rm g}\Sigma_{\rm g}^{\rm all}\frac{\partial}{\partial r}  \biggl(\frac{\Sigma_{\rm g}^{\rm vol}}{\Sigma_{\rm g}^{\rm all}}\biggr)\biggr)= - \dot{\Sigma}^{\rm vol}_{\rm d},  \label{eq:sigma-vol}
\end{align}
where $D_{\rm g}$ is the diffusivity of gas given by Eq.~(\ref{eq:Dg}), 
 $\v_{\rm g}$ is the radial advection velocity of the gas, and 
$ \dot{\Sigma}^{\rm vol}_{\rm d}$ represents the phase change (sublimation/condensation) rate of solid icy particles. The latter two will be described below.
In Equation (\ref{eq:sigma-vol}), the total surface density of gaseous components is 
$\Sigma_{\rm g}^{\rm all} =\Sigma_{\rm g}^{\rm sil} + \Sigma_{\rm g}^{\rm vol}$, where time-dependent $\Sigma_{\rm g}^{\rm sil}$ and $\Sigma_{\rm g}^{\rm vol}$ are calculated from Eqs.~(\ref{eq:sigma-gas1}) and (\ref{eq:sigma-vol}), respectively.

The back-reaction forces that the silicate particles collectively work on the gas modify the gas advection velocity as follows:
In the vertical thickness within the angular momentum of dust particles can be transmitted, $|z|<H_{\rm E}$ (see Sec.~\ref{subsec:d-drift}), the radial gas velocity taking into account both back-reaction of dust and gas viscosity is formulated as \citep{I&G2016,S&O2017, Hyodo+2021}
\begin{align}
\v_{\rm g, BK} = 
 \frac{2\mathrm{St} Z \,\eta \,\v_{\rm K} }{\mathrm{St}^{2} + (1+ Z)^{2}} + 
\left[ 1 - \frac{Z(1+Z)}{\mathrm{St}^{2} + (1+ Z)^{2}} \right] \v_{\rm g, \nu},
\label{eq:vg-BK}
\end{align}
where $\mathrm{St}$ is the Stokes number of silicate solid particles (see the definition in sec.~\ref{subsec:d-drift}) and $\eta$ is given by Eq.~(\ref{eq:eta}).
The first term in the above equation represents the outward flow of the gas resulting from the gain of the angular momentum that the silicate particles lose.
In Eq.~(\ref{eq:vg-BK}), while the previous works on protoplanetary discs have adopted the mid-plane dust-to-gas density ratio, we apply dust-to-gas density within $H_{\rm E}$ defined by Eq.~(\ref{eq:Z}) to correspond to our gas-drag drift formula of dust particles in Eq.~(\ref{eq:v-gas-drag}).
In the upper layers above $H_{\rm E}$, as the gas may not be affected by the dust back-reaction forces, it would advect with unperturbed velocity, $\v_{\rm g, \nu}$ (Eq.~\ref{eq:v-gas}).
Therefore, vertically averaging the above velocities, 
\begin{align}
\v_{\rm g} = \frac{\v_{\rm g, BK} H_{\rm E} + \v_{\rm g, \nu} (H_{\rm g} -H_{\rm E})}{H_{\rm g}}.  \label{eq:vg-w-BKR}
\end{align}
We adopt the same $\v_{\rm g}$ for the silicate vapor and the volatile vapor assuming they are well mixed in a sufficiently short time scale.
We caution that the approximation of $H_{\rm E}$ by the thickness of the Ekman layer is uncertain (see Sec.~\ref{subsec:d-drift}). Our adopted $H_{\rm E}$ value results in much stronger effects of back-reaction of dust to gas than in \citet{Metzger+2012}, as described in Fig.~\ref{fig:wv-fiducial-sigma}a and Sec.~\ref{subsec:ten-to-one}.
Exact estimation of the vertical thickness of the gas layer interacting with dust particles, $H_{\rm E}$, is left for future work.

Assuming water vapor for a representative of volatile vapor, we adopt the molecular weight, $\mu_{\rm vol} = 18 m_{\rm H}$ for the volatile vapor in this paper, where $m_{\rm H}$ is the proton mass.
Inside the silicate sublimation line, silicate vapor of $\mu_{\rm sil} = 30 m_{\rm H}$ is added to the water vapor disc.
This means the mean molecular weight $\mu$ of the gas disc increases from that of the pure volatile vapor disc.
In such two-component mixed gas, we calculate $\mu$ as
\begin{align}
\frac{1}{\mu} =\frac{f_{\rm vol}}{\mu_{\rm vol}} + \frac{f_{\rm sil}}{\mu_{\rm sil}},
\end{align}
where $f_{\rm vol}$ is the mass fraction of volatile vapor in the total gaseous components, and $f_{\rm sil}$ is the mass fraction of silicate vapor.
The addition of $\mu_{\rm sil}$ slightly changes the isothermal sound speed $c_{\rm s}$, which is relevant in determining the parameters of the gas disc and solid particles (see Sec.~\ref{sec:model}).
Although the variation of $\mu$ has little impact on our results, we include the $\mu$-variation for the consistency of the calculations.

After volatile vapor transports outward due to viscous diffusion and gas-solid friction, it would condense into icy particles beyond the snow line.
In this study, we do not model the nucleation of ice solid particles but initially distribute a small enough ice solid particle surface density, $\Sigma_{\rm d}^{\rm vol}$, as seeds of condensation for simplicity.
Following the condensation calculation of silicate vapor in sec.~\ref{subsec:g-d-change}, we determine the surface density change rate of solid icy particles with $\rho_{\rm int}^{\rm vol} =1\ {\rm g \ cm^{-3}}$ as
\begin{align}
\dot{\Sigma}^{\rm vol}_{\rm d} = (R_{\rm con}^{\rm vol}\Sigma_{\rm g}^{\rm vol} - R_{\rm eva}^{\rm vol})\Sigma^{\rm vol}_{\rm d},
\end{align}
where $R_{\rm con}^{\rm vol}$ and $R_{\rm eva}^{\rm vol}$ are
 \begin{align}
R_{\rm con}^{\rm vol} & = 2\sqrt{\frac{k_{\rm B}T}{\mu_{\rm vol}}} \frac{r^{2}_{\rm d}}{m_{\rm d} H_{\rm g}}, \\
R_{\rm eva}^{\rm vol} & = 2\sqrt{2\pi} \frac{r^{2}_{\rm d}}{m_{\rm d}} \sqrt{\frac{\mu_{\rm vol}}{k_{\rm B}T}} P_{\rm eq}^{\rm vol},
\end{align}
and $P_{\rm eq}^{\rm vol}$ is the saturation pressure for the volatile vapor.
Assuming water vapor, we take values of $\mathcal{A}= 6062\ [K]$ and $\mathcal{B}= 30.1$ \citep{Lichtenegger+1991} for $P_{\rm eq}^{\rm vol}$ (Eq.~\ref{eq:P-sat}).

In this study, to highlight the condensation effect of volatile vapor outside the snow line, we neglect the advection/diffusion and sublimation of icy particles.
In other words, we fix the icy particles produced by the condensation at the location where the condensation occurs.
This might be the case due to the ineffective gas-drag drift of icy particles under the significantly small gas density outside the snow line (e.g., Fig.~\ref{fig:wv-fiducial-sigma}a) and due to the slow PR-drag in proportion to the distance from the central star (Eq.~\ref{eq:Mdot_PR_thin} or Eq.~\ref{eq:PR-limit}). 
Modeling of nucleation and the evolution of icy solid particles is left for future work.

 \subsubsection{Simulation settings for silicate disc with volatile vapor}
For the simulation in Sec.~\ref{sec:evo-w-vol}, we adopt the same numerical methods described in Sec.~\ref{subsec:setting} except for the calculated region.
To include the snow line which could be located at $\sim 30 R_{\odot}$ (see below) in the calculated region, 
we take the region from the WD surface ($r=R_{\rm WD}\sim 0.01 \ R_\odot$) out to a radius safely beyond the snow line ($r \sim 100 \ R_\odot$).
The number of radial grids is 700--1000, depending on calculation settings.

Same as in Sec.~\ref{subsec:setting},
we adopt the ring-like distribution 
of the silicate solid disc surface density, $\Sigma_{\rm d}^{\rm sil}$, given by
Eq.~(\ref{eq:initial-sigma-d}) as an initial condition.
In this section, we fix $r_{0} \simeq 0.5  R_{\odot}$ and $\Delta r \simeq 0.2 R_{\odot}$ in Eq.~(\ref{eq:initial-sigma-d}) for all simulations, so that the most of the initial ring mass is located between the silicate sublimation line ($r \simeq 0.1 R_{\odot}$) and the Roche limit radius of rocky materials ($r=1.0  r_{\rm R} \sim 1.3 \ R_\odot$).
We vary $\Sigma_{\rm d0}$ to change the initial silicate disc mass, $M_{\rm disc }^{\rm sil}$; for example,  $\Sigma_{\rm d0} \sim 2$ g/cm$^{2}$ provides $M_{\rm disc}^{\rm sil} \sim 10^{22}$ g, which is the same mass as in Sec.~\ref{sec:res-1comp}.

The icy materials may undergo tidal disruption earlier than the rocky materials because the Roche limit radius of icy materials is located slightly farther than that of rocky materials due to the lower density of ice, $\rho_{\rm int} =1$ g cm$^{-3}$ (see Eq.~\ref{eq:roche}).
After the icy debris is produced, they sublimate to viscously spread to inward and outward. 
As an initial condition, we thus assume the volatile vapor is distributed from inside the Roche limit of rocky materials to outside the Roche limit of icy materials 
($0.5 R_{\odot} < r^{\rm vol} < 2.5 R_{\odot}$) with a surface density profile as
\begin{equation}
\Sigma_{\rm g}^{\rm vol} = \Sigma_{\rm g0}^{\rm vol} \left( \frac{r}{r_{0}} \right)^{-1}.
\label{eq:initial-sigma-g-vol}
\end{equation}
We change the initial disc mass of volatile vapor, $M_{\rm disc}^{\rm vol}$, by changing $\Sigma_{\rm g}^{\rm vol}$.
The surface density with $\Sigma_{\rm g0} \sim 0.3$ g/cm$^{2}$ produces $M_{\rm disc}^{\rm vol} \sim 10^{22}$ g.

In this section, we assume that all particles initially have $r_{\rm d}$ = 1cm, as the observations of dust thermal emission suggested \citep{Graham1990}.
We take $\alpha=10^{-3}$ as in Sec.~\ref{subsec:setting} and fix the vertical thickness ratio of $H_{\rm E}/H_{\rm g}$ at $f=0.1$ (see sec.~\ref{subsec:d-drift}).

\subsection{Volatile-rich Discs: $M_{\rm disc}^{\rm sil}/M_{\rm disc}^{\rm vol} = 1$}
\label{subsec:one-to-one}
In this section, we present the evolution of silicate discs with co-evolving volatile vapor in the case of $M_{\rm disc}^{\rm sil}/M_{\rm disc}^{\rm vol}= 1$.
We then show the resulting accretion rate and time-integrated accretion mass of silicate/volatile components on the WD stellar surface.
Such volatile-rich discs might originate from the disruption of icy bodies which resided in cold outer regions such as Kuiper-Belt analogs.

\begin{figure*}
\hspace{-15.0mm}
\begin{minipage}{0.95\linewidth}
\includegraphics[bb=0 0 756 288, scale=0.67]{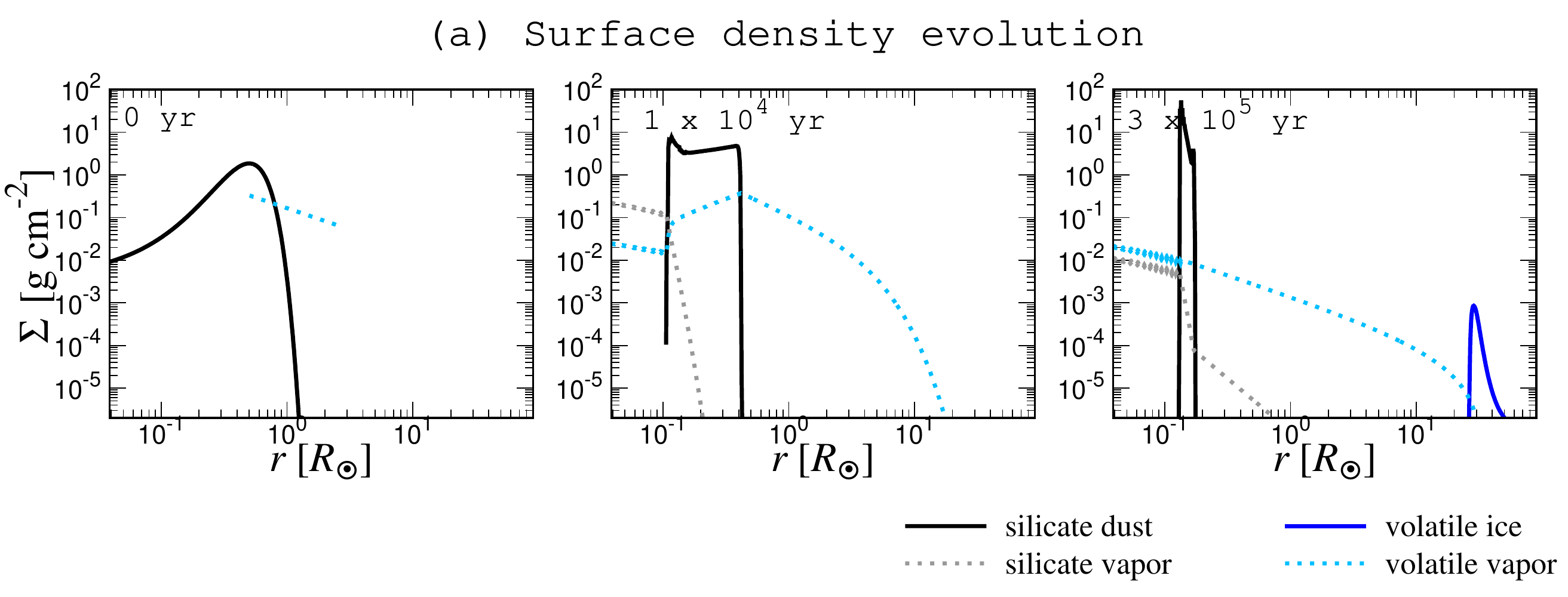}
\end{minipage} 
\\
\begin{minipage}{0.45\linewidth}
\vspace{4.0mm}
\hspace{-4mm}
\includegraphics[bb=0 0 360 252, scale=0.7]{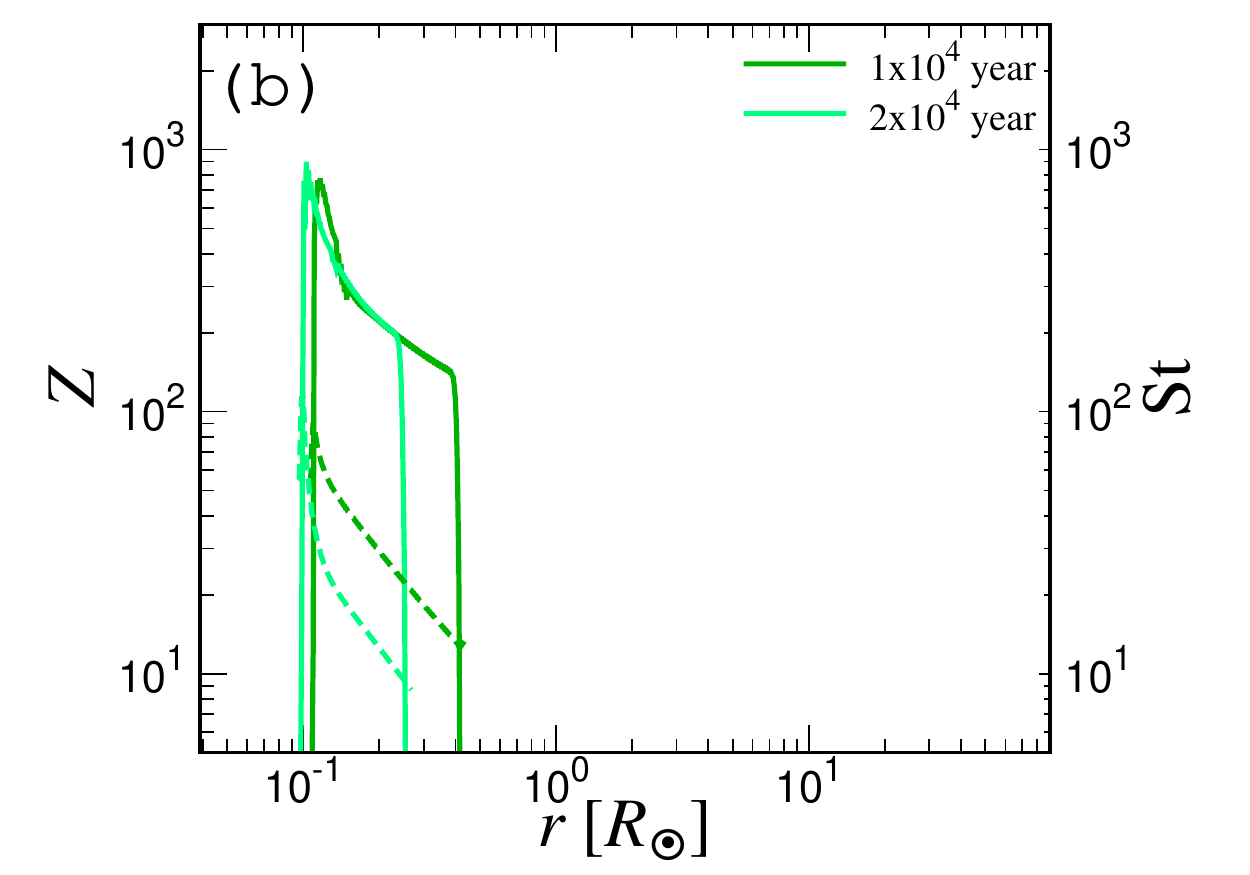}
\end{minipage} 
\hspace{2mm}
\begin{minipage}{0.45\linewidth}
\vspace{4.0mm}
\includegraphics[bb=0 0 360 252, scale=0.7]{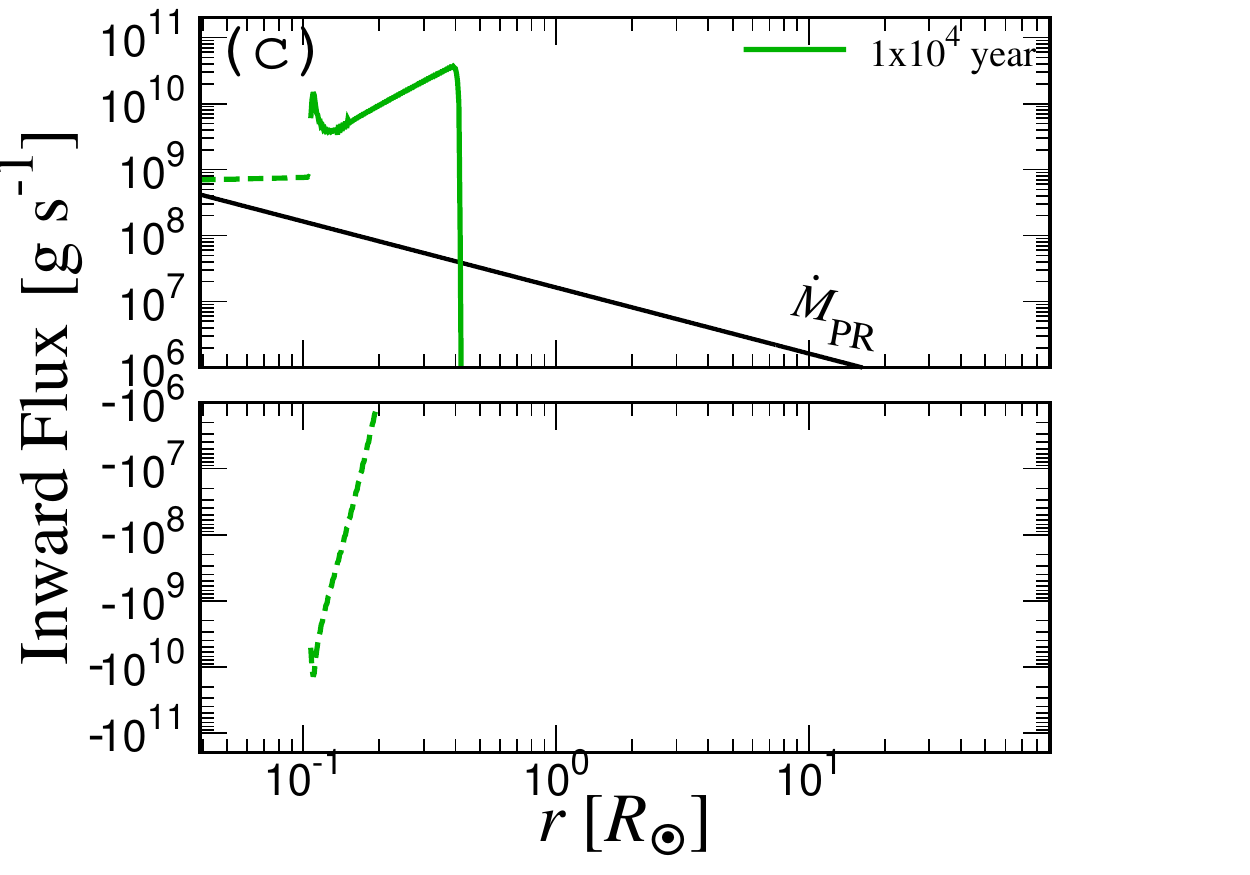}
\end{minipage} 
\caption{
Time evolution of (a) surface density profiles, (b) solid concentration $Z$ (solid lines) and $\mathrm{St}$ (dashed lines), and (c) inward flux of silicate particles (solid line) and vapor (dashed line) obtained from our two-component disc model calculation for the fiducial run.
In top panels, silicate particle surface density $\Sigma_{\rm d}^{\rm sil}$ (black lines), silicate vapor surface density $\Sigma_{\rm g}^{\rm sil}$ (gray dotted lines), icy particle surface density $\Sigma_{\rm d}^{\rm vol}$ (blue lines), volatile vapor surface density $\Sigma_{\rm g}^{\rm vol}$ (light-blue dotted lines), are presented.
In panel (b), the different colors represent physical quantities at different times. 
In panel (c), the black line shows analytically estimated $\dot{M}_{\rm PR,thick}$ (Eq.~(\ref{eq:PR-limit})).
\label{fig:wv-fiducial-sigma}}
\end{figure*}

Taking the case of $M_{\rm disc}^{\rm vol}=10^{22}$ g as an example, the physical quantities that characterize the two-component disc structure are shown in Figure~\ref{fig:wv-fiducial-sigma}.
Outside the silicate line, the volatile vapor discs viscously spread to the snow line $r_{\rm snow}\sim 26 R_{\odot}$ without undergoing condensation. 
It drastically enhances the total gaseous surface density in a region overlapping with silicate dust particles compared to that of the one-component silicate discs (Fig.~\ref{fig:all-w-cond}a).
This provides several orders of magnitude smaller $Z$ (Fig.~\ref{fig:wv-fiducial-sigma}b) than one-component silicate discs, allowing for $\v_{\rm GD} > \v_{\rm PR}$.
In consequence, the mass flux of silicate particles significantly exceeds the limit of PR-drag accretion flux, $\dot{M}_{\rm PR,thick}$, everywhere (Fig.~\ref{fig:wv-fiducial-sigma}c).
The accretion flux of silicate vapor is lower than that of silicate particles, but it is still larger than $\dot{M}_{\rm PR,thick}$.
This decrease is due to the fact that a limited fraction of the particles produced by the re-condensation of diffused-out silicate vapor can drift back to the silicate line (see Sec.~\ref{subsec:with-con} for the description of outward flow structure).

On the other hand, the solid silicate disc affects the evolution of the volatile vapor disc.
Due to the back-reaction of silicate dust particles, the surface density distribution of volatile vapor is much lower than that determined by the viscous evolution alone, $\Sigma_{\rm g} \propto r^{-1}$, in a region overlapping with silicate dust particles (the middle panel of Fig.~\ref{fig:wv-fiducial-sigma}a).
Compared to the results of \citet{Metzger+2012}, that region exhibits much more significant change in the radial profile of vapor disc, and this may come from our assumption on efficiency of the back-reaction, that is, $H_{\rm E}$.
In that region, the surface density ratio of dust-to-gas within $H_{\rm E}$, $Z$, is larger than $\mathrm{St}$, and the value of $Z\sim 100$ tends to be maintained throughout evolution (Fig.~\ref{fig:wv-fiducial-sigma}b).
This is one of the common features when the back-reaction of particles governs two-component disc evolution. Furthermore, the reduction of the volatile vapor accretion rate occurs when $Z>\mathrm{St}$ (see below).

The upper left panel of Figure~\ref{fig:acc-one-to-one} shows the resulting instantaneous accretion rate onto the stellar surface in the case of $M_{\rm disc}^{\rm vol} = 10^{22}$ g. 
In contrast to the mono-compositional silicate disc evolution, high silicate accretion rate of $\dot{M}_{\rm acc}^{\rm sil} \gg \dot{M}_{\rm PR,thick}$ is produced due to gas drag from volatile vapor.
As $Z>\mathrm{St}$, the accretion rate of silicate particles can be estimated by the gas-drag driven flux of silicate particles (Eq.~(\ref{eq:M_GD}) with $q=1$),
\begin{align}
\dot{M}_{\rm GD}  & \simeq  \frac{11}{4} \left(\frac{1}{Z}\right)\ \, f\mathrm{St} \, \v_{\rm K}   \left(\frac{H_{\rm g}}{r}\right)^2 2\pi r\Sigma_{\rm g}  \nonumber \\
& \sim 6 \times 10^{11}\ \frac{\mathrm{St}}{Z}\ \left(\frac{\Sigma_{\rm g}^{\rm all}}{1\ {\rm g\ cm^{-2}}}\right)\ \left(\frac{r_{\rm sil}}{R_{\odot}} \right)\ {\rm g/s}, \label{eq:dotM_GD-est}
\end{align}
where we use $Z=\Sigma_{\rm d}/f \Sigma_{\rm g}$, $f=0.1$, and $r_{\rm sil}/R_{\odot} \sim 0.1$.
Substituting numerically obtained $\mathrm{St} \sim 50 $, $Z \sim 500$, and $\Sigma_{\rm g}^{\rm all} \sim 0.1 {\rm g\ cm^{-2}}$ at $10^{4}$ year in Eq.~(\ref{eq:dotM_GD-est}), $\dot{M}_{\rm GD} \sim 6\times 10^{9}\ {\rm g/s}$. Taking into account an order of magnitude decrease in the vapor flux due to the recycling flow structure at the silicate line, the above estimation well explains the value of $\dot{M}_{\rm acc,sil}$ in the simulation (Fig.~\ref{fig:acc-one-to-one}).
After $10^4$ year, $\dot{M}_{\rm acc}^{\rm sil}$ increases with time to reach a peak value.
Because the back-reaction becomes weak as the region where solid disc exists become narrow, $\Sigma_{\rm g}^{\rm vol}$ increases up to the value determined by viscous diffusion.
Based on Eq.~(\ref{eq:dotM_GD-est}), larger $\Sigma_{\rm g}^{\rm vol}$ results in larger $\dot{M}^{\rm sil}_{\rm acc}$.
%

As the initial disc mass increases, the accretion rate of silicate particles increases (the upper panels of Figure \ref{fig:acc-one-to-one}).
This dependence is never seen in one-component silicate discs, where the accretion rate is always limited to $\dot{M}_{\rm PR}$ in the equilibrium state, even if we increase the disc mass.
The dependence of $\dot{M}_{\rm acc}^{\rm sil} \propto M_{\rm disc}^{\rm vol}$ can be explained as follows:
The larger initial volatile mass provides larger $\Sigma_{\rm g}$.
On the other hand, for $\mathrm{St}$ in the Stokes regime, $\mathrm{St}$ does not depend on $\Sigma_{\rm g}$ (Eq.~\ref{eq:St}).
In addition, we find that the value of $Z$ is usually converged to $\sim 500$ near the silicate line, even if we vary the disc mass but fix $M_{\rm disc}^{\rm sil}/M_{\rm disc}^{\rm vol}$.
Thus, $\dot{M}_{\rm acc}^{\rm sil} \simeq \dot{M}_{\rm GD}$ monotonically increases with the increase of $\Sigma_{\rm g}$ (Eq.~\ref{eq:dotM_GD-est}).
Based on this dependence, the wide range of observationally inferred high accretion rates can be covered by changing the initial mass of the volatile vapor disc.

When the higher silicate accretion rate is produced at $t<5 \times 10^{4}$ years, the accretion rate of volatile, $\dot{M}_{\rm acc}^{\rm vol}$, is always suppressed at one order of magnitude lower value than $\dot{M}_{\rm acc}^{\rm sil}$.
This is because the back-reaction forces from dust particles decrease the surface density of volatile vapor in the region overlapping the silicate particles and reduces the volatile accretion rate.
This suppression occurs when $Z> \mathrm{St}$ is fulfilled everywhere in the silicate particle disc, and the degree of suppression would potentially depend on the efficiency of angular momentum exchange, i.e., $H_{\rm E}$ in our model.
However, after almost all silicate particles accrete onto the star, the back-reaction forces no longer work, resulting in the volatile-rich accretion.
This latter accretion phase would be inconsistent with the observed photospheric compositional trend that the volatile is depleted.
If the snow line location is sufficiently close (e.g., $\sim 3 R_{\odot}$), the depletion timing of volatile vapor is faster, and the volatile-rich accretion phase would be possibly shorter (see Appendix \ref{app:rsnow-dependence}).

As a result of the volatile-rich accretion, the silicate-to-volatile ratio in the total accretion mass approaches to $M_{\rm disc}^{\rm sil} /M_{\rm disc}^{\rm vol}$, that is, one.
The combining effects of condensation of volatile vapor diffused beyond the snow line and the back-reaction of dust to gas reduce the accretion mass of volatile components, but the reduction is only 30 \% of the initial mass.
The reduction mass fraction due to each effect is semi-analytically derived in Appendix \ref{app:volatile-reduction}.
The above evolution trend and $M_{\rm acc}^{\rm sil} /M_{\rm acc}^{\rm vol}$ is independent of the disc mass as long as $M_{\rm disc}^{\rm sil} /M_{\rm disc}^{\rm vol}$ is fixed.
Therefore, we could conclude that the discs with $M_{\rm disc}^{\rm sil} /M_{\rm disc}^{\rm vol}=1$ could not reproduce the silicate-rich composition in the almost all accretion phase ($> 5 \times 10^{4}$ year) for short $t_{\rm sink}$ nor long $t_{\rm sink}$.

\begin{figure*}
\includegraphics[bb= 0 0 720 540,scale=0.65]{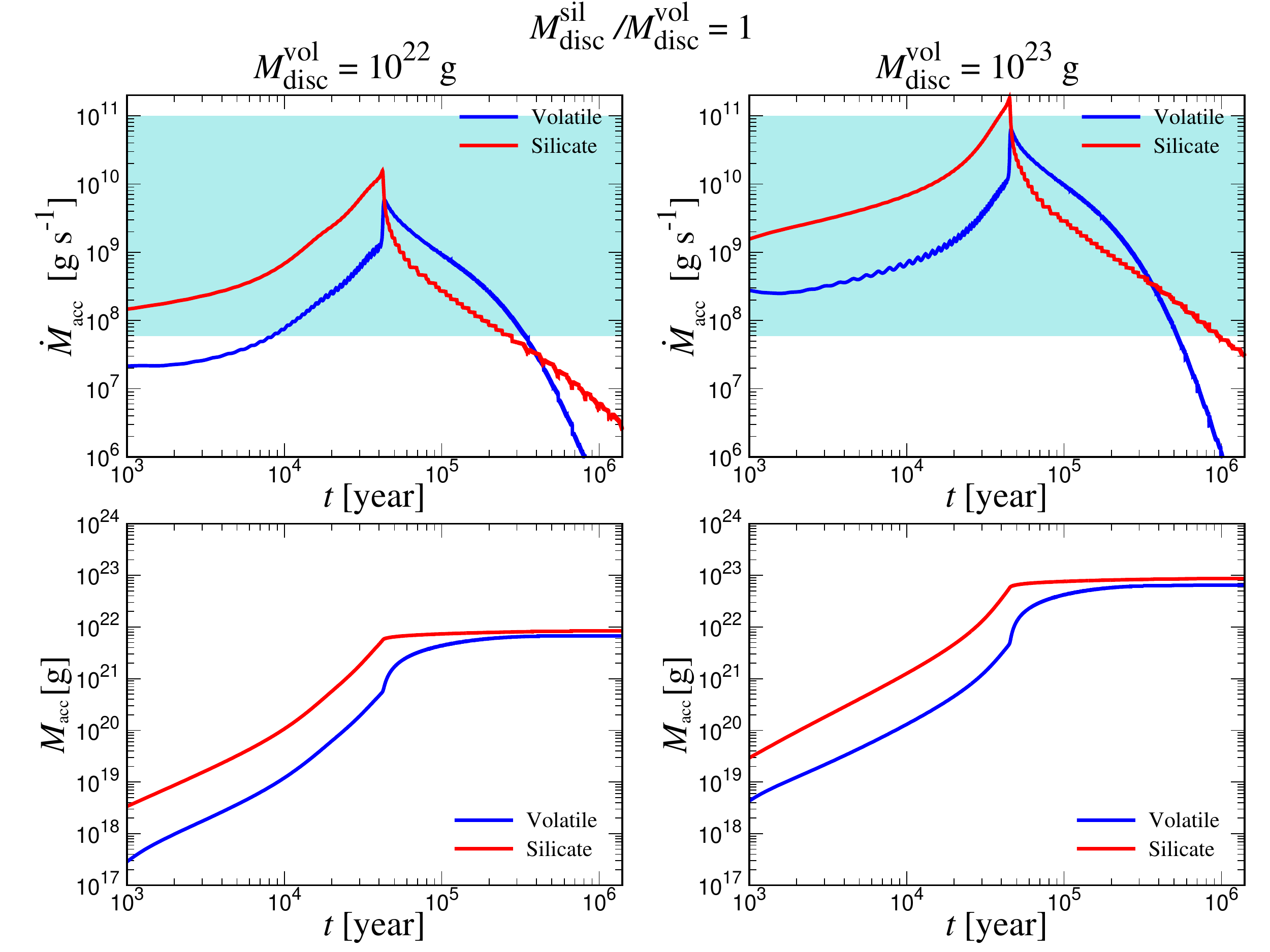}
\caption{
Time evolution of the accretion rate (upper panels) and the cumulative accretion mass (lower panels).
The left and right panels show the results for the disc mass of $10^{22}$ g and $10^{23}$ g, respectively.
The red lines represent the silicate component, and the blue lines represent the volatile component.
In the upper panels, the cyan region highlight the observed values $\dot{M}_{\rm acc,obs}$ higher than $\dot{M}_{\rm PR,thick}$ evaluated at the inner boundary of optically thick region (Eq.~(\ref{eq:PR-limit})).
\label{fig:acc-one-to-one}}
\end{figure*}

\subsection{Discs with Less Volatile Vapor: $M_{\rm disc}^{\rm sil}/M_{\rm disc}^{\rm vol} = 10$}
\label{subsec:ten-to-one}
In section \ref{subsec:one-to-one}, for a fixed volatile and silicate ratio at unity, we have found that the silicate particles predominantly fall with a high accretion rate initially, but it is followed by an icy-rich accretion phase.
In this section, we start with the disc with larger silicate-to-volatile mass ratio, $M_{\rm disc}^{\rm sil}/M_{\rm disc}^{\rm vol} = 10$, to investigate the possibility of avoiding an icy-rich accretion phase.
The origin of such discs might correspond to the C-type asteroid analogs, which are mainly composed of rocky materials but include a small fraction of water ice.

\begin{figure*}
\includegraphics[bb= 0 0 720 540, scale=0.65]{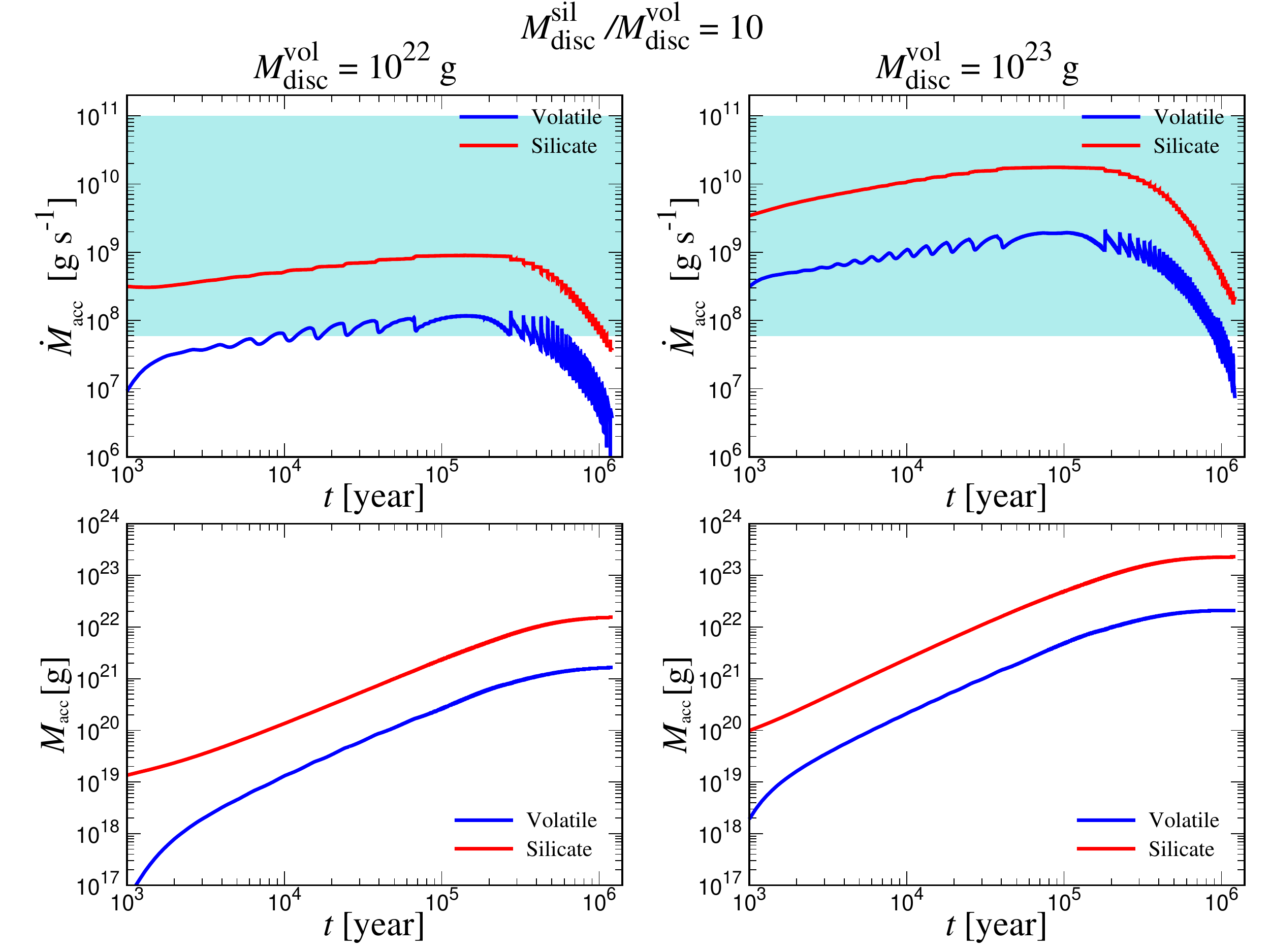}
\caption{
Same as Figure~\ref{fig:acc-one-to-one} except that the silicate-to-volatile ratio in the initial disc mass is 10.
\label{fig:acc-ten-to-one}}
\end{figure*}

Figure~\ref{fig:acc-ten-to-one} shows the
instantaneous accretion rate (upper panels) and the accretion mass (lower panels).
Even for the silicate disc mass significantly larger than the volatile disc mass, the gas drag drift from volatile vapor is effective, and $\dot{M}_{\rm acc}^{\rm sil}$ is able to exceed $\dot{M}_{\rm PR, thick}$.
Furthermore, $\dot{M}_{\rm acc}^{\rm sil}$ increase with the increase of the disc mass, as in the case of $M_{\rm disc}^{\rm sil}/M_{\rm disc}^{\rm vol} =1$.
Comparing the results for the same initial mass of volatile vapor discs between $M_{\rm disc}^{\rm sil}/M_{\rm disc}^{\rm vol} =1$ and $10$, $\dot{M}_{\rm acc}^{\rm sil}$ is somewhat smaller for larger $M_{\rm disc}^{\rm sil}/M_{\rm disc}^{\rm vol}$. 
Because larger $M_{\rm disc}^{\rm sil}/M_{\rm disc}^{\rm vol}$ provides larger $Z$, the back-reaction slows down the silicate accretion Eq.~(\ref{eq:dotM_GD-est}).
As $M_{\rm disc}^{\rm sil}/M_{\rm disc}^{\rm vol}$ increases further, $\dot{M}_{\rm acc}^{\rm sil}$ would asymptote that of one-component silicate discs, and the dependence on the disc mass would vanish.

On the other hand, the evolution track is distinct from the case of $M_{\rm disc}^{\rm sil}/M_{\rm disc}^{\rm vol} =1$.
For the larger $M_{\rm disc}^{\rm sil}/M_{\rm disc}^{\rm vol}$ case,
as the volatile vapor disc is completely exhausted before the silicate solid disc accretes, 
silicate-rich composition of the accretion flux,  $\dot{M}_{\rm acc}^{\rm sil}/\dot{M}_{\rm acc}^{\rm sil}\sim10$ is retained over all phases of the accretion.
This allows for a high silicate-to-volatile ratio of $\sim 10$ even in the total accretion mass (the bottom panels of \ref{fig:acc-ten-to-one}).
Note that the ratio of 10 does not result from the silicate-to-volatile ratio in the initial disc mass but the back-reaction effects of silicate particles to volatile vapor.
In addition, the minor difference between the cases of $M_{\rm disc}^{\rm sil}/M_{\rm disc}^{\rm vol} =1$ and 10 is the presence/absence of the increase in $\dot{M}_{\rm acc}^{\rm sil}$ with time.
Because the back-reaction force exert on the volatile vapor until the end of the volatile vapor disc accretion for $M_{\rm disc}^{\rm sil}/M_{\rm disc}^{\rm vol} =10$, $\Sigma_{\rm g}^{\rm vol}$ keeps a low constant value.
Without the elevation of $\Sigma_{\rm g}^{\rm vol}$, $\dot{M}_{\rm acc}^{\rm sil}$ is almost unchanged (Eq.~\ref{eq:dotM_GD-est}).
%

In addition, the back-reaction of silicate particles prolongs the lifetime of volatile vapor comparing to the disc evolve under the viscous accretion alone.
This behavior cannot seen in \citet{Metzger+2012}, possibly resulting from the much stronger effects of back-reaction of dust to gas than in \citet{Metzger+2012} (see Sec.~\ref{subsec:vol-model}).
With our settings of $\alpha=10^{-3}$ and $r_{\rm disc} \sim 1.5R_{\odot}$ (the initial location of the volatile disc), the viscous time-scale of gas disc of volatile would be $r_{\rm disc}^2/3\nu \sim 2 \times 10^4$ year. 
On the other hand, the lifetime of volatile vapor reaches  $\sim 10^6$ years because of the slow down of viscous accretion due to the back-reaction (Fig.~\ref{fig:acc-ten-to-one}).
As the back-reaction is likely to be stronger for larger $M_{\rm disc}^{\rm sil}/M_{\rm disc}^{\rm vol}$, lifetimes of volatile vapor are longer in the case of $M_{\rm disc}^{\rm sil}/M_{\rm disc}^{\rm vol}$=10. This allows the presence of volatile vapors to continue to affect the accretion rate of refractory solids for an extended period of time.

In conclusion, the coupling to volatile vapor could simultaneously provide high-$\dot{M}$ and refractory-rich photospheric composition when $M_{\rm disc}^{\rm sil}/M_{\rm disc}^{\rm vol} =10$.
The refractory-rich composition would be attained for any metal-sinking timescale $t_{\rm sink}$, and hence both DA-type and DB-type WDs.
However, the volatiles is mixed to the accretion onto the stellar photosphere in 10\% by mass. If observational constraints allow the 10\% inclusion, our results may possibly lead to a solution to the high-$\dot{M}$ puzzle. 
We compare this fraction in more detail to the observationally inferred fraction of volatiles in Sec.~\ref{subsec:s-to-v-ratio}.

\section{Discussion}
\label{sec:discussion}
In Sec.~\ref{sec:res-1comp} and Sec.~\ref{sec:evo-w-vol}, we focus on the reproduction of two observational trends; higher accretion rate than the PR drag flux and the rocky-rich photospheric composition.
In this section, not limited to these two points, we discuss the observation facts that can and cannot be explained by our current disc model.

\subsection{Co-existence of Solid and Vapor Discs}
\label{subsec:co-exist}
For one-component silicate discs, the effect of re-condensation in our model is to drastically reduce the vapor density outside the silicate sublimation radius, $r_{\rm sil}$.
This is because the timescale of re-condensation is several orders of magnitude shorter than that of outward advection/diffusion of silicate vapor.
Even if we decrease the effectiveness of re-condensation by reducing the solid surface area, which the vapor can interact with, to the plate surface area (Eq.~\ref{eq:C-cond}), these results hardly change.

This immediate removal of silicate vapor could reconcile with observations that the gaseous emission lines have been detected in only a small number of systems \citep{Manser+2020}. 
However, turning to such gaseous systems, there is a potential problem with our proposed model: Most of the gaseous systems exhibit the refractory solid and refractory vapor discs spatially co-exist over a large range of radii, which seems to contradict our model suggesting effective removal of vapor.
Based on the doppler-shift of gaseous emission lines due to the Keplerian motion, gas discs are inferred to extend to $\sim 1 R_{\odot}$, and this approximately coincides with the outer edge of solid discs \citep{Melis+2010}.
Moreover, the emission pattern in the gas disc of SDSS J1228 mainly comes from the outside of 0.6 $R_{\odot}$ \citep{Manser+2016}, whereas the disc of solids extends much closer in \citep{Brinkworth+2009}.
Motivated by these observations, previous disc studies \citep{Rafikov2011b, Metzger+2012} have theoretically modeled overlapping silicate solid and vapor without the effects of condensation.

One possibility to solve this issue might be the the photo-ionization heating by ultraviolet radiation.
Due to this heating mechanism, the temperature of the gas in the disc upper layer, which is optically thin for host star's direct UV irradiaon, could be much higher than the blackbody equilibrium temperature predicted in Eq.~(\ref{eq:T-prof}) \citep{Melis+2010}.
The observed gas is actually as hot as $\sim 5000$ K \citep[e.g.,][]{Melis+2010, Hartmann+2016}.
Therefore, the condensation of silicate vapor would not occur in the disc upper layer, and therefore the gas may exist even if beyond the mid-plane silicate line.

As the scale height of gas with $\sim 5000$ K is several orders of magnitude larger than the vertical thickness of dust particle discs, they would be vertically decoupled \citep{Melis+2010}.
To verify such decoupling, it would be needed to assess whether it takes a longer time for the upper gas to be mixed down toward the mid-plane or not. We will leave it for future work. 
In addition, if the upper hot gas layer and dust particles are decoupled, the runaway accretion driven by gas-drag drift acceleration is unlikely to occur.

%

\subsection{Volatile Mass Fraction}
\label{subsec:s-to-v-ratio}
For the silicate discs including volatile vapor, the effects of the back-reaction of dust particles to volatile vapor in our model provide a silicate-rich composition to both accretion flux and accumulated accretion mass, but they contain at least $\sim 10\%$ of volatiles by mass (Sec.~\ref{subsec:ten-to-one}).
On the observation side, we have not yet reached a high enough accuracy to measure volatile content of 10\% even if present in the WD atmospheres, and therefore it may be premature to conclude whether our results are consistent or not.
Although the discussion would be tentative, in this section, we compare the volatile mass fractions obtained in our simulations with those constrained from the photospheric observations so far.
In addition, we address the issue that volatiles are rarely detected in the emission lines from the discs.

As hydrogen never gravitationally sinks in WDs with helium-dominated atmospheres (DB-type), trace hydrogen in DB-type WDs is thought to potentially reflect the water accretion history \citep{Jura+2009, Gentile-Fusillo+2017}.
\citet{Jura+2012} assembled 57 DB-type WDs within 80 pc of the Sun and estimated that their summed H accretion rate is $<1.4 \times 10^{7}$ g/s while the summed heavier atoms accretion rate is $1.6 \times 10^{10}$ g/s. 
This indicates that the hydrogen would account for $<0.1$ wt\% of the total mass accretion rate.
On the other hand, assuming $\sim 10$ wt\% of volatile vapor as H$_2$O vapor, our simulation results predict that hydrogen comprise $10$ wt\% $\times\ 2/18 \sim 1$ wt\% of accreting mass.
This is at least one order of magnitude larger than the observationally constrained H fraction, 0.1 wt\%, although the \citet{Jura+2012}'s sample included the low accretion rate systems, which might be mono-compositional silicate discs.

To explain the observational value, the silicate-to-volatile mass ratio in the accretion mass would have to be larger than 10.
This may be possible if the angular momentum exchange between dust and gas is more efficient. It accelerates the gas-drag drift of silicate particles while moving more volatile-vapor mass outward.
As a result, silicate-to-volatile mass ratio in the accretion mass could be larger than 10 even if we start with $M_{\rm disc}^{\rm sil}/M_{\rm disc}^{\rm vol}=10$.
As the vertical turbulence of gas would determine the efficiency of angular momentum exchange through $H_{\rm E}$, larger $\alpha_z> 10^{-3}$ might make such a situation possible (Eq.~\ref{eq:f}).
As another possibility, for the WDs with high temperature, the radiation pressure driven by large EUV luminosity, in particular, Lyman-$\alpha$ line may inhibit the accretion of H atoms \citep{Gansicke+2019}, which would be produced by the dissociation of water vapor.  
The strong depletion of H compared to other atoms has been reported in the gas disc only composed of volatile elements \citep{Gansicke+2019}.

Oxygen in excess of expected for oxide metals would be another evidence of water accretion \citep{Farihi+2013, Raddi+2015, Farihi+2016, Hoskin+2020}.
Although there is a handful of systems indicating oxygen excess, we summarize the inferred water mass fraction of all the systems in Table \ref{tab:acc-water}.
The water fraction is close to 10 wt\% for two systems, while the other systems show a much larger water mass fraction than 10 wt\%.
The former may correspond to the accretion with $M_{\rm disc}^{\rm sil}/M_{\rm disc}^{\rm vol}=10$ (Sec.~\ref{subsec:ten-to-one}), and the latter might correspond to the volatile-rich accretion periods which takes place in the accretion with $M_{\rm disc}^{\rm sil}/M_{\rm disc}^{\rm vol}$=1 (Sec.~\ref{subsec:one-to-one}).

Furthermore, all of the water accreting systems tend to show a metal accretion rate higher than that produced by PR drag, $\dot{M}_{\rm PR} \sim 10^{8}$ g/s, and some of their values are higher than $\dot{M}_{\rm PR}$ by orders of magnitude.
Table \ref{tab:acc-water} collects the total accretion rate of each system, which is observationally inferred.
This might imply the potential association of water accretion with the high metal accretion rate, as our simulation results expect.

\begin{table}
 \caption{$\dot{M}_{\rm acc}$ and estimated water mass fraction}
 \label{tab:acc-water}
 \begin{tabular}{lccc}
  \hline
  WD Name & SpT$^{\rm a}$ & H$_2$O mass & total $\dot{M}_{\rm acc}$  \\
          &  & wt\% & g/s \\
  \hline
  GD 61$^{(1)}$  & DBAZ & 26 & 2.7$\times 10^{8}$\\
  SDSS J1242+5226$^{(2)}$  & DBAZ & 38 & 2.0 $\times 10^{10}$\\
  WD 1536+520$^{(3)}$   & DBAZ & 5-10 & $ 4.2 \times 10^{9} $ \\
  WD J2047-1259$^{(4)}$ & DBAZ & 8$^{\rm\ b}$ & $8.8 \times {10^{8}}^{\rm\ b}$\\
  \hline 
 \end{tabular}
 {\begin{flushleft}
 \footnotesize{
 $^{\rm a}$ In SpT (Spectral type of the WD), DBAZ means the He-dominant atomosphere with trace H and metals. \\
 $^{\rm b}$ The values are derived assuming steady state phase. \\
 References (1) \citet{Farihi+2013}; (2) \citet{Raddi+2015}; (3) \citet{Farihi+2016}; (4) \citet{Hoskin+2020} }
 \end{flushleft} }
\end{table}

Other than water ice, volatile elements such as N, C, and S could form solid ice in planetary bodies.
However, these elements are unlikely to be responsible for the volatile vapor that promotes gas-drag drift of silicate particles:
The detection of these elements is not so common, and the measured mass ratio or upper limit of C/Si is an order of $10^{-2}-10^{-3}$ \citep{Gansicke+2012}.
If we assume volatile vapor as carbon-monoxide and silicate as forsterite for example, C/Si in our simulation would be $\sim 0.2$, which is several orders of magnitude larger than the observational ratio.
%

For the disc of volatile vapor, there is a gap between our proposed calculations and the observations on disc emission lines:
Our disc simulations predict that the surface density of volatile vapor is substantially larger than that of silicate gas in the region overlapping with the silicate particle discs (Fig.~\ref{fig:wv-fiducial-sigma}a).
In contrast, whereas the emission lines of metallic vapor have been detected in that region in several systems (see section \ref{subsec:co-exist}), those of volatile gas such as H and O are observed 
in only one system \citep{Gentile-Fusillo+2021}.
This contradiction cannot be solved in the current framework of the proposed calculations, but it would be important to reveal whether the lack of the volatile emission line comes from the absence of volatile vapor itself or the disc properties such as temperatures and geometries and/or observational wavelength ranges \citep[e.g.,][]{Hartmann+2016,Steele+2021}.
The assessment by combining our disc evolution simulation with photo-chemical and radiative-transfer calculations is left for future work.


\section{Conclusion} \label{sec:conclusion}

We have revisited the observed high accretion rate of compact planetary debris discs onto white dwarf (WD) stars that cannot be attained by the mass flux, $\dot{M}_{\rm PR,thick}$, by the Poynting-Robertson (PR) drag alone \citep{Rafikov2011}.
We have also investigated the possibility that such high-$\dot{M}$ is produced by the accretion with refractory-rich composition, suggested by observations on polluted atmospheres \citep[e.g.,][]{Jura&Young2014}.
\citet{Metzger+2012} developed the first accretion disc model that formulates interaction between silicate particles and silicate vapor to propose that the mono-compositional refractory disc could reproduce the high accretion rate by runaway silicate solid accretion caused by increasing silicate vapor.
Although \citet{Rafikov2011b, Metzger+2012} pointed out the potential importance of re-condensation of the silicate vapor,  their model did not incorporate the condensation effect, considering the observations suggesting co-existence of silicate solids and vapors. 

In order to address this problem, we have performed one-dimensional advection/diffusion simulations of silicate dust particles and vapor with their sublimation/re-condensation in a self-consistent manner.
For this, we have applied the method developed by \citet{Hyodo+2019, Hyodo+2021}, which formulates the gas drag as a force exerting on individual particles but takes into account the particle collective effects by including the back-reaction of solid to gas.
This assumption would be different from \citet{Metzger+2012} that 
treated the gas drag as a plate drag. However, when the back-reaction is significantly strong (i.e., $Z>1$ and $Z>\mathrm{St}$), the drag-driven mass accretion rate derived in this work and in \citet{Metzger+2012} shares the same dependence on the gas surface density, which is essential for the runaway accretion.

We have found that in the one-component silicate disc, condensation entirely inhibits the runaway accretion proposed by previous works \citep{Rafikov2011b, Metzger+2012}.
In both cases where we adopt our gas-drag drift formula derived from the particle-based approach and the plate-based approach of \citet{Metzger+2012}, the runaway accretion is completely halted by condensation (Fig.~\ref{fig:acc_rate-w-cond} and Fig.~\ref{fig:acc_rate-w-cond-plate}).
The silicate vapor density outside its sublimation line is solely determined by the saturating vapor pressure
because the condensation timescale is short enough compared with the advection and diffusion.
The radial dependence of the saturation pressure is so strong
that co-existence of solid particles and vapor is limited.
The environment of solid-dominated ($Z \gg 1$) and weak coupling between dust and vapor ($\mathrm{St}\gg 1$), both of which diminish the runaway accretion, is established (Fig.~\ref{fig:all-w-cond}), resulting in a steady accretion state supplied by the mass flux due to the PR-drag.
Therefore, the accretion rate produced from the one-component silicate disc is exactly equal to $\dot{M}_{\rm PR,thick}$.
The rate is independent of the disc parameters such as particle size and solid disc mass.
For the observationally suggested co-existence of silicate solids and vapor, we have raised the possibility that photo-ionized hot gas layer exists in the optically thin upper region, pointing out that such a possible gas layer does not revive the runaway accretion caused near the disc midplane (Sec.~\ref{subsec:co-exist}).

We have additionally performed simulations coupling vapor of volatile elements (e.g., water vapor) that is produced by the sublimation of icy materials included in the infalling planetary bodies.
As the dynamical arguments imply the supply of planetary bodies from the wide orbital regions \citep{Mustill+2018, Li+2022}, their volatile content may be diverse.
In the case that the silicate-to-volatile mass ratio in the initial disc, $M_{\rm disc}^{\rm sil}/M_{\rm disc}^{\rm vol}$ is 1, the silicate accretion rate can be higher than $\dot{M}_{\rm PR,thick}$ for higher disc mass of volatile vapor (Fig.~\ref{fig:acc-one-to-one}).
However, the silicate-rich accretion period is always followed by the volatile-rich accretion period, and the volatile fraction in the total accretion mass reaches up to $\sim 50$\%. This conflicts with the observations on photospheric composition.

In contrast, for $M_{\rm disc}^{\rm sil}/M_{\rm disc}^{\rm vol}=10$, the suppression of volatile vapor accretion due to the back-reaction continues until the end of the volatile disc lifetime (Fig.~\ref{fig:acc-ten-to-one}).
This maintains a silicate-rich composition throughout the whole accretion phase. Although the larger dust-to-gas density decreases the efficiency of accretion of solids, the silicate accretion rate  is still higher than $\dot{M}_{\rm PR,thick}$ due to gas drag and increases with the increase of the disc mass of volatile vapor.
The C-type asteroid analogs like Vesta are a plausible origin of the discs with small but non-negligible fraction of volatiles.

Coupling to volatile might be a possible clue to the high-$\dot{M}$ puzzle although the compositional issue is not completely solved.
With the assumed efficiency of angular momentum exchange between dust and gas, our simulation results predict that the fraction of volatile in total accretion mass is 10 wt\%.
Such inclusion could not be dismissed with the current precision of the photospheric observation while the observations so far might suggest the much lower contents in accretion flux.
As WD ultraviolet luminosity may possibly prevent the accretion of H atoms around the WDs of high temperatures \citep{Gansicke+2019}, it might be an important next step to incorporate the effects of WD radiation into our accretion disc evolution model.
Furthermore, it will enable us to predict the emission lines from the gas discs and therefore to approach the remained issue that our currently proposed calculation could not consistently explain the lack of volatile emission lines from most of the discs.

\section*{Acknowledgements}
The authors thank the referee for insightful comments, which greatly improve our manuscripts.
We also thank Dimitri Veras, Masahiro Ikoma, Ted von Hippel, Hiroshi Kobayashi, and Yuka Fujii for useful discussions, and Boris G\"{a}nsicke for helpful comments on the revision. This work was supported by JSPS KAKENHI Grant Number JP22J00632 / 21J13248 / 21H04512
and MEXT Kakenhi Grant 18H05438. R.H. acknowledges the financial support of MEXT/JSPS KAKENHI (Grant Number JP22K14091). R.H. also acknowledges JAXA's International Top Young program.

\section*{Data Availability}
The data underlying this article will be shared on reasonable request to the corresponding author.







\appendix

\section{Runaway accretion calculated from the Metzger et al. (2012) model} \label{app:M12}
To examine the influence of the gas-drag treatment on the runaway accretion, we here simulate the disc evolution without condensation, assuming the drift formula of the plate-based model (Eqs.~\ref{eq:M_GD-M12} and \ref{eq:vGD-M12}).
This is the reproduction calculation of runaway accretion presented in \citet{Metzger+2012}, and obtained results are directly compared to those calculated from our model that assumes gas drag on individual particles but takes into account the dust back-reaction in Sec.~\ref{subsec:M12}.
For this, we set our model in Sec.~\ref{sec:model} with the tuning by
i) turning off the condensation, and
ii) using $\v_{\rm GD,M12}$ (Eq.~\ref{eq:vGD-M12}) that is obtained from \citet{Metzger+2012}'s drag force per unit surface area, but
iii) maintaining a weak enough particle diffusion (our code does not allow zero diffusion). 
For a fiducial value of $\alpha=10^{-3}$, we choose $Re_{\ast}=0.3$ so that the combination of $\alpha$ and $Re_{\ast}$ satisfies a condition for the runaway accretion that the feedback parameter safely exceeds 1  \citep[see][for more details]{Metzger+2012}. The initial particle radius is 0.1 cm.

\begin{figure}
\hspace{-2.5mm}
\includegraphics[bb= 0 0 360 252, width=\columnwidth]{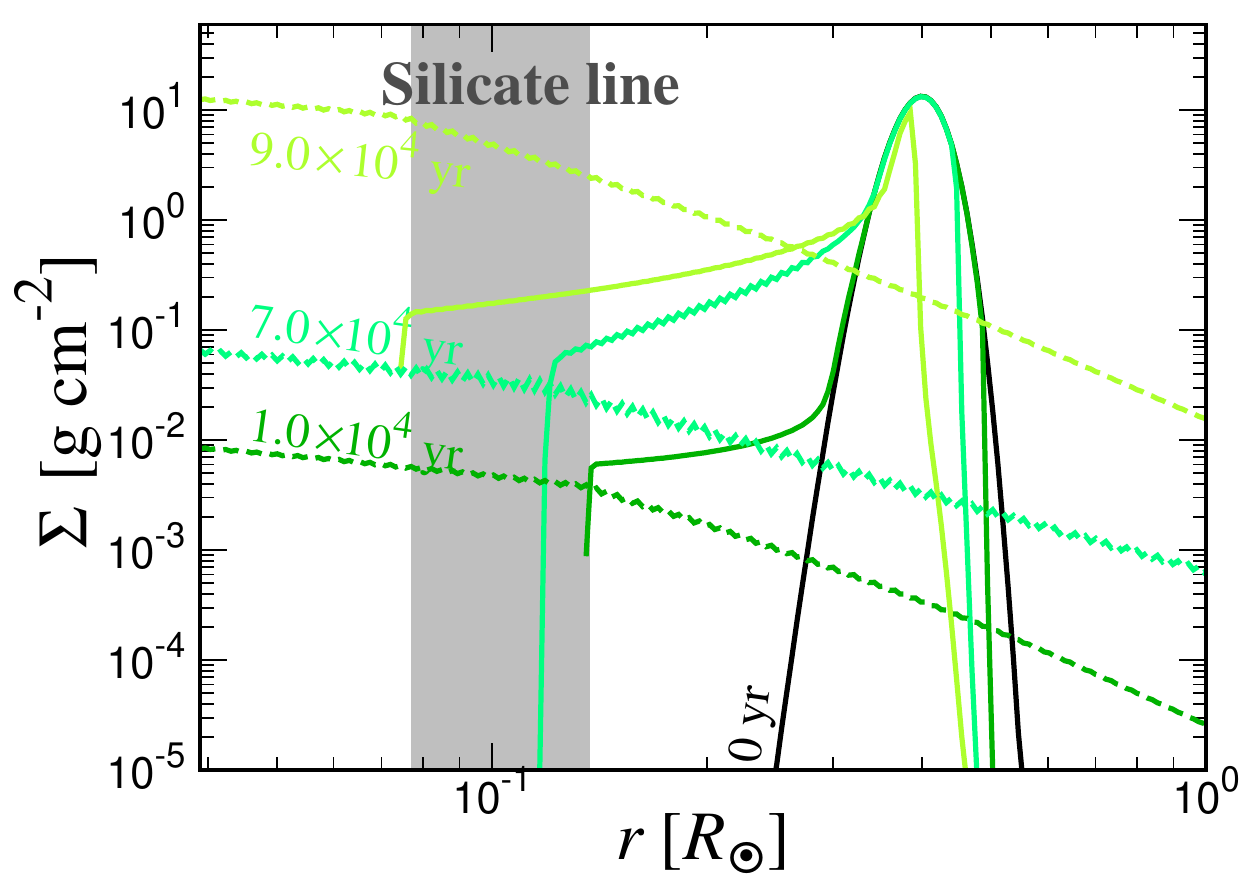}
\vspace{-1mm}
\caption{Same as Figure \ref{fig:sigma-wo-cond} except that the calculation was done using gas-drag drift derived from the plate drag regime, $\v_{\rm GD}$, in Eq.~(\ref{eq:vGD-M12}).
}
\label{fig:sigma-wo-cond-plate}
\end{figure}

\begin{figure}
\hspace{-2.5mm}
\includegraphics[bb= 0 0 360 252, width=\columnwidth]{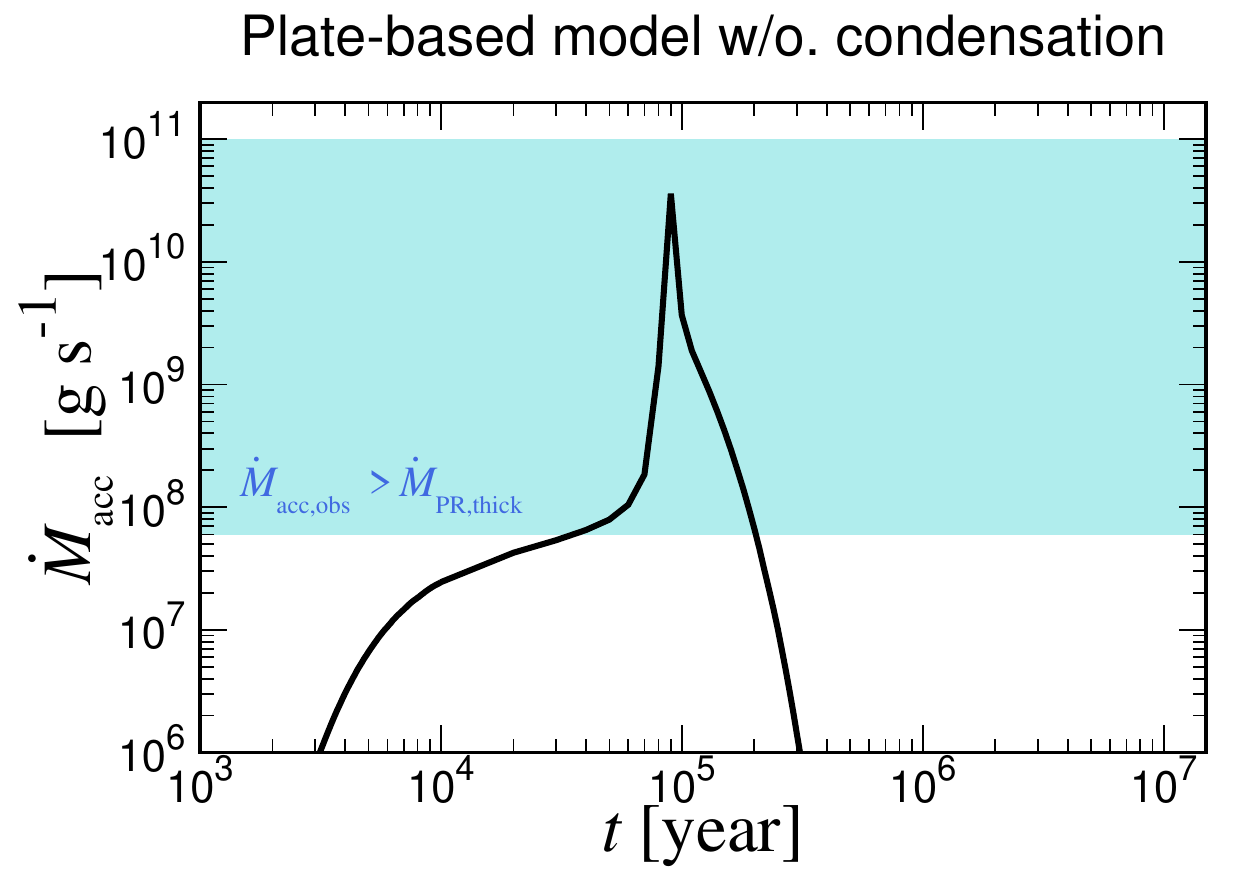}
\vspace{-1mm}
\caption{
Same as Figure \ref{fig:acc_rate-wo-cond} except except that the calculation was done using  $\v_{\rm GD}$ in Eq.~(\ref{eq:vGD-M12}).
\label{fig:acc_rate-wo-cond-plate}}
\end{figure}

The obtained surface density evolution and the accretion rate evolution onto WDs are shown in Figures \ref{fig:sigma-wo-cond-plate} and \ref{fig:acc_rate-wo-cond-plate}, respectively.
Compared to the results calculated from our particle-based model in Sec.~\ref{subsec:M12}, the difference of gas-drag drift velocity of dust particles only quantitatively changes the dust surface density distribution in the last phase of runaway accretion  (see Figs.~\ref{fig:sigma-wo-cond-plate} and \ref{fig:sigma-wo-cond}).
Although the disc parameter choice affects the accretion rate evolution track,
we can conclude that the runaway accretion phenomenon occurs commonly in two models (see Sec.~\ref{subsec:M12} for more details).

\section{Semi-analytical estimation of accretion mass of volatiles}\label{app:volatile-reduction}
\begin{figure}
\hspace{-2.5mm}
\includegraphics[bb= 0 0 360 252, width=\columnwidth]{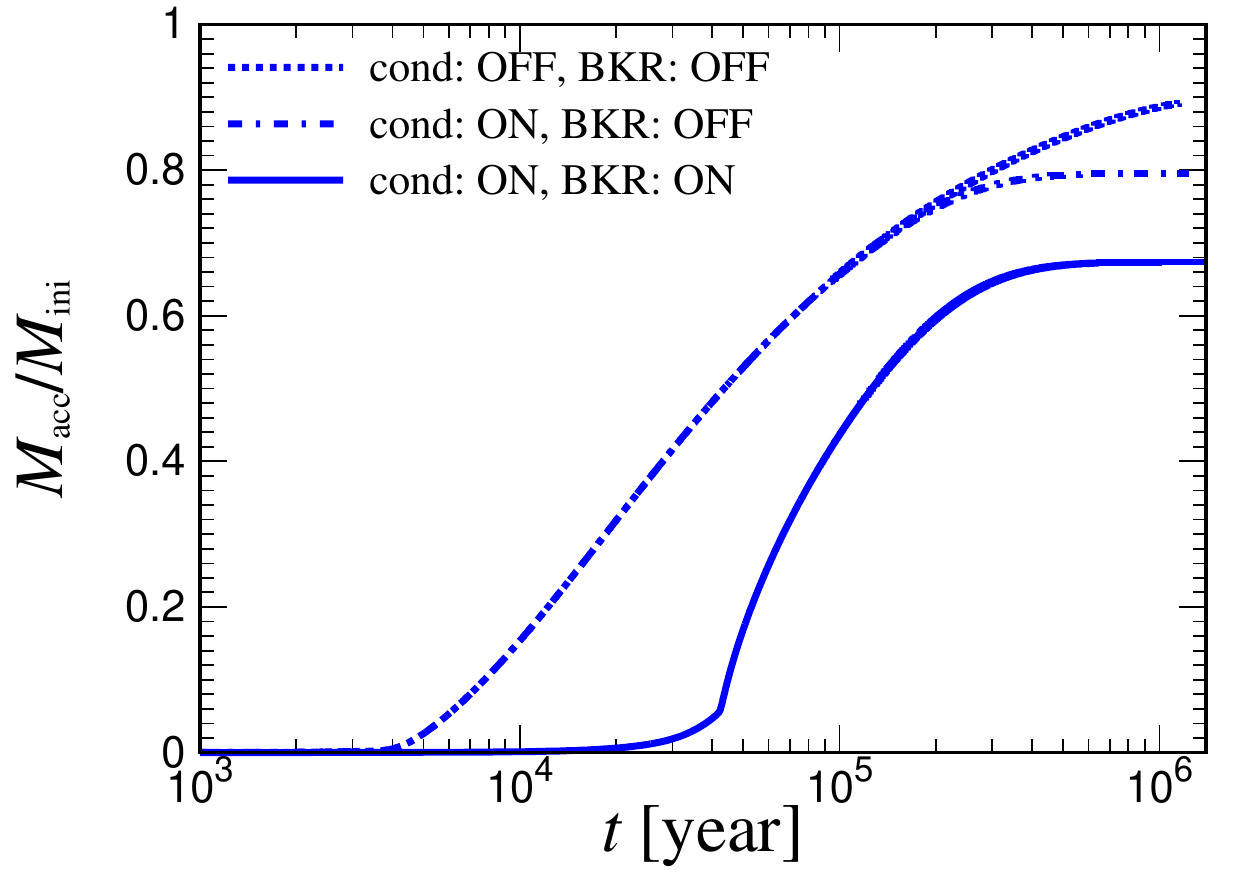}
\vspace{-1mm}
\caption{
Time evolution of total accretion mass of volatile scaled by its initial disc mass, $10^{22}$ g.
The dotted line shows the result turning off the condensation of volatile vapor diffused beyond the snow line and back-reaction of dust particles to gas.
The dotted-dashed line shows the result removing only the back-reaction of dust particles to gas. 
\label{fig:acc_mass-w-vol-linear}}
\end{figure}

We here quantify the both effects of condensation of volatile vapor diffused beyond the snow line and the back-reaction of dust to gas on the final accretion mass of volatiles onto the WD surface, $M_{\rm acc}^{\rm vol}$.
By turning on or off each effect, we demonstrate how much each of these two effects reduces $M_{\rm acc}^{\rm vol}$ in Figure ~\ref{fig:acc_mass-w-vol-linear}.
Without the above two effects, almost all the initial volatile vapor disc mass $M_{\rm disc}^{\rm vol}$ accretes onto the WD surface until $10^6$ year.
If turning on the condensation of volatile vapor only, $M_{\rm acc}^{\rm vol}$ is $\sim 0.8 M_{\rm disc}^{\rm vol}$. The addition of back-reaction of dust to gas further reduces $M_{\rm acc}^{\rm vol}$ to $\sim 0.67 M_{\rm disc}^{\rm vol}$.

The final accreting mass of the volatile vapor, $M_{\rm acc}^{\rm vol}$, can be analytically estimated by applying the discussion for the debris disc and moon formation around proto-Earth via giant impact \citep{Ida+1997}.
The mass conservation and angular momentum conservation give
\begin{align}
L_{\rm disc}^{\rm vol} + \Delta L_{\rm GD} = M_{\rm acc}^{\rm vol} \sqrt{R_{{\rm WD}}} + (M_{\rm disc}^{\rm vol}-M_{\rm acc}^{\rm vol} ) \sqrt{r_{\rm snow}} \label{eq:am-vol},
\end{align}
where  $L_{\rm disc}^{\rm vol}$ is the initial angular momentum of volatile vapor disc initially located at $r_{\rm disc}^{\rm vol}$, $\Delta L_{\rm GD}$ is the gain in the angular momentum due to solid-gas friction, $R_{\rm WD} = 0.01 R_{\odot}$, and $r_{\rm snow}$ is the location of the snow line (sublimation line of the volatile vapor). 
Using the initial angular momentum of silicate solid disc, $L_{\rm disc}^{\rm sil}$, located at $r_{\rm disc}^{\rm sil}$,  $\Delta L_{\rm GD}$ is expressed as $\xi_{\rm GD}L_{\rm disc}^{\rm sil} =  \xi_{\rm GD}M_{\rm disc}^{\rm sil} \sqrt{r_{\rm disc}^{\rm sil}}$ where  $\xi_{\rm GD}$
is the loss fraction of the silicate dust angular momentum by gas drag.
In Eq.~(\ref{eq:am-vol}), we set $GM_{\rm WD} = 1$.
By dividing $\sqrt{R_{\odot}}$, Eq.~(\ref{eq:am-vol}) can be rewritten as 
\begin{align}
M_{\rm disc}^{\rm vol} \left[ 1 + \xi_{\rm GD} \left( \frac{M_{\rm disc}^{\rm sil} }{M_{\rm disc}^{\rm vol} } \right)
\left( \frac{r_{\rm disc}^{\rm sil} }{r_{\rm disc}^{\rm vol} } \right)^{1/2} \right] \left( \frac{r_{\rm disc}^{\rm vol} }{R_{\odot} } \right)^{1/2} \nonumber \\ 
= 0.1 M_{\rm acc}^{\rm vol} + C^{1/2}_{\rm snow} (M_{\rm disc}^{\rm vol}-M_{\rm acc}^{\rm vol} ),
\end{align}
where $C_{\rm snow} = r_{\rm snow}/R_{\odot}$.
Assuming $C_{\rm snow}^{1/2} \gg 0.1$, this leads to the final volatile accreting mass,
\begin{align}
M_{\rm acc}^{\rm vol}  \simeq \left \{ 1 - C^{-1/2}_{\rm snow} \left[ 1 + \xi_{\rm GD} \left( \frac{M_{\rm disc}^{\rm sil} }{M_{\rm disc}^{\rm vol} } \right)
 \left( \frac{r_{\rm disc}^{\rm sil} }{r_{\rm disc}^{\rm vol} } \right)^{1/2} \right]  \left( \frac{r_{\rm disc}^{\rm vol} }{R_{\odot} } \right)^{1/2} \right \} M_{\rm disc}^{\rm vol}. \label{eq:acc-ice}
\end{align}

If the angular momentum exchange does not occur (i.e., $\xi_{\rm GD}=0$),  
$M_{\rm acc}^{\rm vol}$ becomes $0.80 M_{\rm disc}^{\rm vol}$ for $C_{\rm snow} \sim 26$, where we assume $r_{\rm disc}^{\rm sil} \sim r_{\rm disc}^{\rm vol}/3$ and $r_{\rm disc}^{\rm vol} \sim 1.5 R_{\odot}$. In other words, $0.20 M_{\rm disc}^{\rm vol}$ is re-condensed to remain as an icy ring at $r_{\rm snow}$. 
This well explains $M_{\rm acc}^{\rm vol}$ obtained from the numerical simulation with turning on the condensation but turning off the back-reaction of dust to gas.
Taking into account the angular momentum exchange (i.e., $\xi_{\rm GD}>0$), the volatile vapor reaching the snow line increases.
 If most of silicate particle mass drifts to the silicate sublimation line at $ r_{\rm sil}$ due to gas drag from the volatile vapor, $\xi_{\rm GD} = 1- \sqrt{r_{\rm sil} / r_{\rm disc}^{\rm sil}} \simeq 0.55$, where we assume $r_{\rm disc}^{\rm sil} \sim 5 r_{\rm sil}$.
Substituting the value into Eq.~(\ref{eq:acc-ice}), $M_{\rm acc}^{\rm vol} \simeq 0.68 M_{\rm acc}^{\rm vol}$, and this reproduces the simulation result turning on both the condensation of volatile vapor and the back-reaction of dust to gas. 

\section{Possible impacts of snow line location} \label{app:rsnow-dependence}

We here examine the possible influence of the location of volatile vapor sublimation line, $r_{\rm snow}$. 
The condensation of volatile vapor outside the snow line could reduce the volatile accretion mass onto the star, and the icy ring mass produced by condensation would be larger for smaller $r_{\rm snow}$ (see Appendix \ref{app:volatile-reduction}).
Because silicate dust particles inside the snow line could block the direct stellar irradiation, the disc temperature might be lower than blackbody temperature determined by the direct irradiation from the WD Eq.~(\ref{eq:T-prof}).
This may results in smaller $r_{\rm snow}$ than that obtained in Sec.~\ref{sec:evo-w-vol}.
Accordingly, we vary $r_{\rm snow}$ to repeat the fiducial run with $M_{\rm disc}^{\rm vol} = M_{\rm disc}^{\rm sil} = 10^{22}$ g. 
In the simulation, $r_{\rm snow}$ is changed by changing the coefficient in the saturation pressure, $\mathcal{B}$ in Eq.~(\ref{eq:P-sat}) for simplicity, instead of changing disc temperature structure in Eq.~(\ref{eq:T-prof}).

\begin{figure}
\hspace{-2.5mm}
\includegraphics[bb = 0 0 360 252, width=\columnwidth]{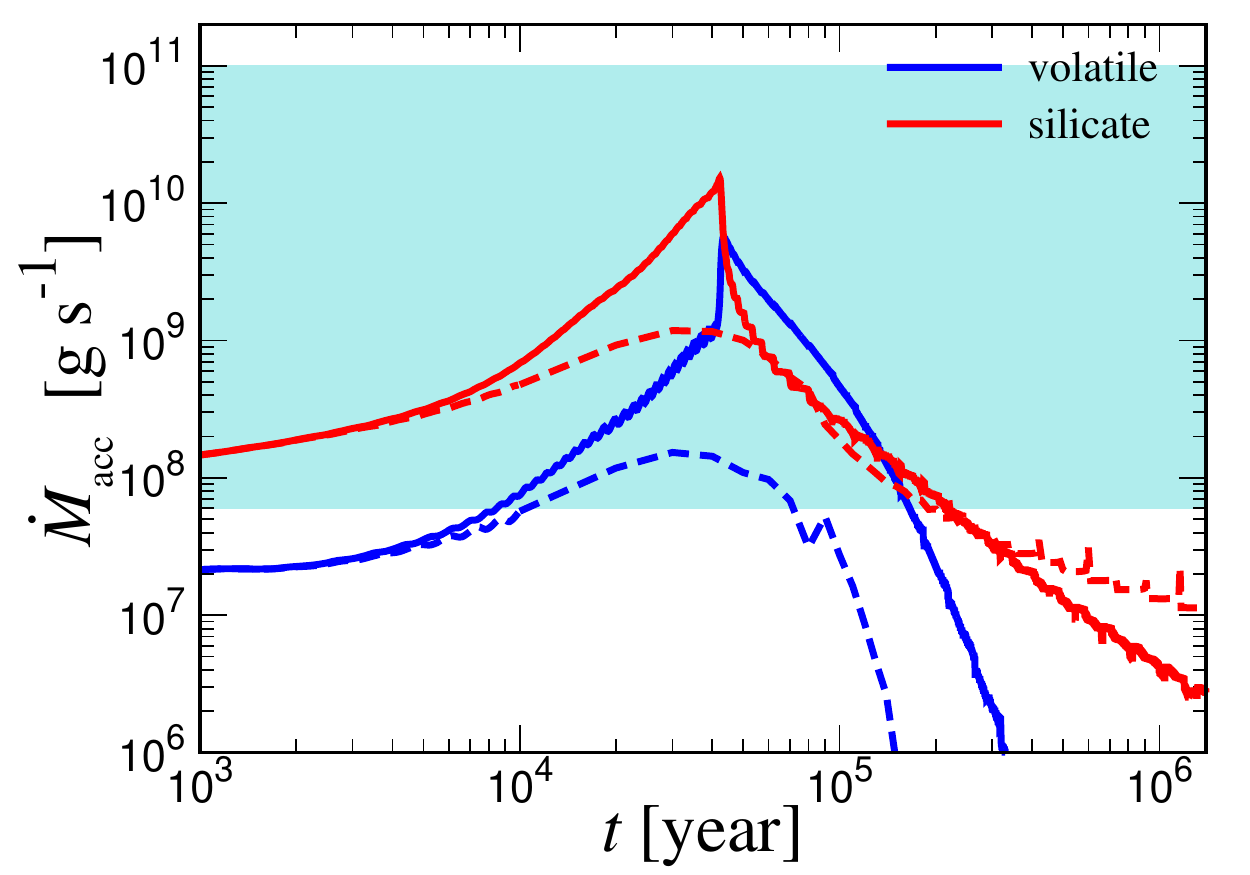}
\vspace{-1mm}
\caption{Same as the upper right panel of Figure \ref{fig:acc-one-to-one} except for $r_{\rm snow} \sim 3 R_{\odot}$ (dotted lines) and $r_{\rm snow} \sim 10 R_{\odot}$ (solid lines).
\label{fig:acc_rate-w-vol-rsnow}}
\end{figure}

Figure~\ref{fig:acc_rate-w-vol-rsnow} presents the accretion rate evolution of silicate and volatile with various assumed $r_{\rm snow}$. 
For smaller $r_{\rm snow}$, $\dot{M}_{\rm acc}{\rm vol}$ drops earlier
because volatile vapor reaches $r_{\rm snow}$ to start condensation earlier.
In particular, for $r_{\rm snow} \sim 3 R_{\odot}$, the volatile vapor is completely consumed before all of the silicate solid materials accrete.
This resembles the evolution path of $M_{\rm disc}^{\rm sil}/M_{\rm disc}^{\rm vol} = 10$ (Fig.~\ref{fig:acc-ten-to-one}), which maintain a constant high $\dot{M}_{\rm acc}^{\rm sil}$ and provide $\dot{M}_{\rm acc}^{\rm sil}/\dot{M}_{\rm acc}^{\rm vol} \sim 10$ at all times.
As a result, the silicate-to-volatile ratio is 10 also in the total accretion mass.
This suggests that if $r_{\rm snow}$ is sufficiently small, even the volatile-rich disc can reproduce the high accretion flux and silicate-rich composition for WDs with any metal-sinking time.
The icy ring mass obtains $\sim 95 \%$ of $M_{\rm disc}^{\rm vol}$ for $r_{\rm snow} \sim 3 R_{\odot}$, in contrast to $\sim 30 \%$ of $M_{\rm disc}^{\rm vol} $ for $r_{\rm snow} \sim 30 R_{\odot}$.

We will leave the full disc calculation self-consistently incorporating the temperature evolution for future work.
In addition, the icy particle advection/diffusion outside the snow line should be included because their inward drift affects the mass of remaining icy rings.

\bsp	
\label{lastpage}
\end{document}